\documentclass[aps,prc,%tightenlines,
     twoside,twocolumn,pre,floatfix]{revtex4-1}
 
\usepackage{tabularx} 
\usepackage{epsfig}
\usepackage{bm}
\usepackage{graphicx}
\usepackage{float}
\usepackage{gensymb}
\usepackage{amsmath} 
\usepackage{amssymb} 
\usepackage{subfig}
\usepackage{tensor}
\usepackage[hidelinks]{hyperref}
\usepackage{wrapfig}
\usepackage{color}
\usepackage{xcolor}
\usepackage{multirow}
\usepackage{hhline}
\colorlet{darkgreen}{green!50!black}
\colorlet{brightyellow}{yellow!75!red}
\colorlet{orange}{red!50!yellow}
\colorlet{darkblue}{blue!60!black}
\colorlet{darkred}{red!80!black}

\newcommand{\comment}[1]{}

\newcommand{\pcal}{\mathcal{P}}

\newcommand{\threej}[6]{%
\begin{pmatrix}
#1 & #2 & #3 \\
#4 & #5 & #6
\end{pmatrix}}

\begin{document}

\title{Quark Coalescence: Formation of Mesons Including Excited States}

\author{R.~J.~Fries}
\affiliation{Cyclotron Institute and Department of Physics and Astronomy, Texas A\&M University, College Station, TX~~77845, USA}

\author{P.\ Virupapuram}
\affiliation{Cyclotron Institute and Department of Physics and Astronomy, Texas A\&M University, College Station, TX~~77845, USA}

\author{J.\ Purcell}
\affiliation{Cyclotron Institute and Department of Physics and Astronomy, Texas A\&M University, College Station, TX~~77845, USA}

\author{H.\ Anconetani}
\affiliation{Cyclotron Institute and Department of Physics and Astronomy, Texas A\&M University, College Station, TX~~77845, USA}

\author{W.\ Lippincott}
\affiliation{Cyclotron Institute and Department of Physics and Astronomy, Texas A\&M University, College Station, TX~~77845, USA}

\author{S.\ Robicheaux}
\affiliation{McNeese State University}
\affiliation{Cyclotron Institute and Department of Physics and Astronomy, Texas A\&M University, College Station, TX~~77845, USA}

\author{M.\ Kordell}
\affiliation{Cyclotron Institute and Department of Physics and Astronomy, Texas A\&M University, College Station, TX~~77845, USA}

\author{C.~M.~Ko}
\affiliation{Cyclotron Institute and Department of Physics and Astronomy, Texas A\&M University, College Station, TX~~77845, USA}

\date{\today}

\begin{abstract}
We discuss the quantum mechanics of coalescence of quark-antiquark pairs into mesons using a non-relativistic quark model. We derive the coalescence probabilities assuming a harmonic oscillator potential and generic Gaussian wave packet shapes for the initial quarks and antiquarks. Our particular emphasis is on modeling excited states of the meson spectrum consistently. We provide the formalism to systematically include excited states from the Particle Data Book, and many more predicted by the quark model, up to $L=4$ and masses of about 2.2 GeV for light flavors. We provide estimates of masses and decay branching ratios for unconfirmed states. We use a phase space picture which is appropriate for the quasi-classical nature of the information typically available for the quarks and antiquarks in applications like Monte Carlo simulations. We demonstrate that for typical parton configurations expected in jets, excited meson states are populated abundantly.
\end{abstract}

\maketitle

\section{Introduction}
\label{sec:1}

Hadronization has been a challenging feature of Quantum Chromodynamics (QCD) since its inception \cite{Gross:2022hyw}. Bound states involve highly non-perturbative aspects of QCD, including quark confinement. Over the past decades, several models have emerged to describe the phenomenon of hadronization empirically. Among them are string fragmentation \cite{Andersson:1983ia,Sjostrand:2006za}, cluster hadronization \cite{Webber:1983if,Marchesini:1991ch} and statistical hadronization \cite{Becattini:2009fv}. The string and cluster hadronization models, through implementation in sophisticated Monte Carlo event generators like PYTHIA \cite{Sjostrand:2006za,Sjostrand:2014zea} and HERWIG \cite{Marchesini:1991ch,Bellm:2015jjp}, have been able to describe experimental data in great detail for both electron-position ($e^++e^-$) collisions, and for jets and related observables in proton-proton ($p+p$) collisions. Statistical hadronization has had success in describing statistical properties, like average hadron abundances, for a wide range of collision systems \cite{Becattini:1995if,Becattini:1995if,Andronic:2005yp,Andronic:2017pug}.

Quark coalescence, or quark recombination, had been proposed in the 1970s as an alternative to these models \cite{Das:1977cp,Jones:1979xj,Migneron:1981fw}. Quark recombination assumes that well-defined states of quarks and antiquarks exist prior to hadronization, which become valence quarks in hadrons through coalescence using an explicit attractive interaction between them. This contrasts with the other models in which hadrons or their valence quarks are mostly created at hadronization itself by decay of a gluon field, decay of a large-mass color-neutral cluster, or partition of available energy. There are obvious connections between these models, which we will not touch on in detail in this work. However, it is also obvious that there are substantial differences. Information encoded in the partons before hadronization, e.g.\ about correlations, is more easily transferred onto hadrons in quark coalescence models.
Precisely this feature revived interest in this approach in the early 2000s, when data arrived from high energy nucleus-nucleus ($A+A$) collisions exhibiting some surprising features \cite{Greco:2003mm,Greco:2003xt,Fries:2003vb,Fries:2003kq,Fries:2004ej,Molnar:2003ff}.
%which had been dormant for a while, but was resurrected in .
Quark recombination naturally described enhanced production of baryons vs mesons in nuclear collisions, and it offered an explanation for the constituent quark number scaling seen in the azimuthal modulation of collective motion of hadrons, in particular the valence quark scaling of elliptic flow $v_2$ \cite{Fries:2008hs}. 
The recombination model has since been used profusely to model hadron formation in nuclear collisions \cite{Oh:2009zj,ExHIC:2010gcb,Ravagli:2007xx,Fries:2025jfi}.
The theoretical description was often informed by similar recombination processes observed for nucleons in the formation of light nuclei \cite{Baltz:1995tv,Scheibl:1998tk,Chen:2003ava}.
%which could not reliably be described by other models.

For nuclear collisions, this has led to a paradigm in which calculations for jet and hard hadron observables typically employ perturbative parton showers with string fragmentation. On the other hand, hadron observables with typical momenta between 2 and 10 GeV/$c$, as well as many heavy flavor observables are often computed using a quark recombination picture \cite{Fries:2025jfi}. It is therefore desirable to develop an approach that incorporates aspects of several models, e.g.\ quark recombination and string fragmentation, in a way that is self-consistent and can be applied universally to all collision systems and hadron energies. One such model is Hybrid Hadronization \cite{Han:2016uhh,JETSCAPE:2023ewn,JETSCAPE:2025wjn}, in which a given parton system is checked for direct coalescence between quarks and antiquarks; any remaining partons form strings, which then fragment. This model is then applied to all "final-state" quarks and antiquarks, independent of their origin. The decision which partons are "final state" is outside of the purview of the hadronization model, which takes this information as input. In Hybrid Hadronization, the coalescence mechanism naturally dies off in favor of string fragmentation as partons become more isolated in phase space. The transition is controlled by the number of excited hadron states that are included in the recombination channel. Pre-exisiting information on color flow is preserved for the string fragmentation phase, as only color singlets are formed in coalescence. Ultimately, it was shown that in Hybrid Hadronization quark recombination dominates for large phase space densities of partons, more likely at small observable momenta and in larger collision systems, while string fragmentation dominates for smaller phase space densities, typically seen at larger momenta and in smaller collision systems \cite{JETSCAPE:2025wjn}.

In this work, we focus solely on the quark recombination model, albeit numerical simulations are carried out using Hybrid Hadronization. The Hybrid Hadronization code in the JETSCAPE and X-SCAPE event generators \cite{Jetscape:github,Putschke:2019yrg} labels hadrons according to their production mechanism, making it easy to separate production channels.
Our goal is to provide a simple but comprehensive and self-consistent formalism for production of unpolarized mesons from a given distribution of quarks and antiquarks. Our focus will be on the inclusion of the complete known spectrum of mesons, with the option to add "missing" quark model states that have not been experimentally confirmed \cite{Fries:2023miu,Fries:2024jfw}. A similar work for baryons is in preparation \cite{Virpapuram:2025bar}.

To be precise, we choose the non-relativistic quark model with an attractive and confining potential as the basis of our work. The confining property does not have phenomenological relevance here, as the tower of excited states is truncated for practical reasons, and the confining property would be enforced otherwise, e.g. in Hybrid Hadronization through string dynamics at large distances. The quark-antiquark interaction potential is that of an isotropic harmonic oscillator. This choice is made for simplicity as a Wigner phase space formalism is available and all integrals are analytically solvable \cite{KORDELLII2022168960}. We discuss below why a phase space formalism is preferable for our applications. An interested reader could replace the results of Sec.\ \ref{sec:2} by those of a different potential, e.g.\ a Coulomb or Cornell potential.
Of course, our treatment is only an approximation to the full quantum field theoretical description of hadronization. However, the non-relativistic quark model has been surprisingly successful in predicting the tower of hadron states, as well as masses, magnetic momenta, and other key properties \cite{PhysRevD.58.114017}. We do not wish to use our model to compute such hadron observables to high accuracy. The main application is to make predictions for observables at particle colliders averaged over many events, like momentum spectra of hadrons and correlations of hadrons. Macroscopic features like momentum and energy conservation, the number and degeneracies of states, and their decay patterns, play dominant roles.

Within the approximation, our approach will employ a full quantum mechanical treatment, albeit limited by the input for partons, which typically does not contain their complete quantum information. We include a rather comprehensive set of quark model meson states --- 330 in total, not counting spin degeneracies ---  which covers all states in the particle data book \cite{ParticleDataGroup:2020ssz}, plus additional quark model states up to a cutoff. We will discuss below how masses and decay branching ratios of unconfirmed quark model states are estimated. We will also briefly explore the phenomenological consequences of including highly excited states in some applications at the end of this manuscript, but our focus will be on establishing the necessary framework. 

We summarize the goals of the paper once more in detail. Assume that an ensemble of quarks and antiquarks is given. Flavor, momentum and position of each parton are known, while their spin polarizations are random. Color is taken as random, although some knowledge about color or correlations of color can be implemented as discussed below. This is the typical input provided by shower Monte Carlo codes or Boltzmann transport codes for partons. The spatial information is sometimes absent in codes intended for simple collision systems like $e^++e^-$ or $p+p$, but becomes an integral part of codes in $A+A$ collisions. The information provided is essentiallly classical. To connect to a quantum mechanical picture that makes use of all of the information, we interpret this information in the phase space (Wigner) picture of quantum mechanics.
One now picks one quark-antiquark pair from this ensemble randomly. We will compute the probability for this pair to form meson bound states with given set of quantum numbers. We will not distinguish the different polarization states of the mesons.

The paper is organized as follows. In the next section, we review meson wave functions in the quark model and then proceed to compute the recombination probabilities, first focusing on the spin-flavor part of the wave functions. Subsequently, we compute the spatial wave function overlap, however using Wigner's phase space formalism. In Sec.\ \ref{sec:3}, we discuss how the model is implemented in Hybrid Hadronization, and we introduce the meson multiplets that can be used. In Sec.\ \ref{sec:4}, we present an example of an application in $e^++e^-$ collisions and study some basic features of our formalism.
Finally, we summarize and present an outlook.

\section{Quark-Antiquark Coalescence}
\label{sec:2}

In this section, we compute the coalescence probabilities for a quark-antiquark pair with given momenta at given positions.

\subsection{Meson Wave Functions in the Quark Model}

The wave functions of hadrons in the quark model can generally be built from linear combinations of the form
\begin{equation}
  \Psi_h = \zeta_{\mathbf{1}} \sum_i d_i \, \chi_i \phi_i \psi_i (\mathbf{x}_1,\mathbf{x}_2, \ldots)
\end{equation}
where the $\zeta$, $\chi$, $\phi$, and $\psi$, respectively, are color states, spin states, flavor states, and spatial 
wave functions, respectively, built from the valence quarks of the hadrons with spatial coordinates $\mathbf{x}_1, \mathbf{x}_2, \ldots$. The $d_i$ are coefficients where $i$ symbolically runs over all combinations of quantum numbers allowed for a particular hadron.
The color wave function can always be factored out, as only color singlet wave functions  $\zeta_{\mathbf{1}}$ are allowed.

In the following, we focus on mesons as bound states of a quark and an antiquark. Then, $\zeta_{\mathbf{1}}$ represents the SU(3) singlet in $\mathbf{3}_c \otimes \bar{\mathbf{3}}_c = \mathbf{8}_c \oplus \mathbf{1}_c $, i.e.\
\
\begin{equation}
    \zeta_\mathbf{1} =\frac{1}{\sqrt{3}} \left( \zeta_r      
    \zeta_{\bar r} + \zeta_b \zeta_{\bar b} + \zeta_g 
    \zeta_{\bar g}  \right)
\end{equation}
in terms of  a ($r$,$b$,$g$) color basis.
Likewise, the flavor state can be factored out for most mesons, as it is uniquely determined, except for a few cases of mixing for states of the same quantum numbers. This happens in particular between $u\bar u$, $d\bar d$ and $s\bar s$ states which will be dealt with later on a case-by-case basis. 

Mesons are angular momentum eigenstates that can be classified in spectroscopic notation as $n{}^{2S+1}L_J$ through their total angular momentum $J$, orbital angular momentum $L$ and spin quantum number $S$. Radial excitations add another quantum number $n$. All quantum numbers are integers with $n$ starting at 1, and angular momentum quantum numbers starting at 0. As usual, $L+S\geq J \geq |L-S|$
and the quantum number $L$ is often replaced by the notation S, P, D, F, G, etc., for $L=0,1,2,3,4$, respectively.
The wave function for a meson with polarization $J_z$ can thus be further specified to be 
\begin{equation}
\label{eq:mesongeneral}
  \Psi_M^{J_z} %\left( {n}^{2S+1} L_J \right) 
  = \zeta_{\mathbf{1}} \phi_M 
  H_{M}^{n,L} (x)
  \!\!\!\!\!\!
  \sum_{S_z+L_z=J_z} \!\!\!\!\!\! C^{J, J_z}_{L,L_z;S,S_z}  \chi_{S}^{S_z} Y_{L}^{L_z}   \, ,
\end{equation}
where we have suppressed the angle arguments of the spherical harmonics $Y_{L}^{L_z}$
for brevity. $H_M^{n,L}$ is the radial wave function which depends on $x=|\mathbf{x}_2-\mathbf{x}_1|$.
The $C^{J, J_z}_{L,L_z;S,S_z}$ are Clebsch-Gordan coefficients, and $S_z$, $L_z$, and $J_z$ are the quantum numbers for the $z$-components of their respective angular momentum operator. The spin wave function for each meson is fixed to be either from the $S=1$ triplet or $S=0$ singlet of the SU(2) product of two spin doublets
$\mathbf{2}_S \otimes {\mathbf{2}}_S = \mathbf{3}_S \oplus \mathbf{1}_S $.
In the quark model, the radial wave functions $H_M^{n,L}(x)$ are the only pieces which need to be determined from QCD dynamics. %Let us look at two examples, the $rho+(770)$, a well known ${}^3S_1$ state, and one of its $P$-wave counterparts, the 

\subsection{Recombination Probabilities}

The quantum state of a quark or antiquark $j$ can be written as
\begin{equation}
  \label{eq:qwf}
  \psi^{s_z}_{j,c_j} (\mathbf{x})= \zeta_{c_j} q^{s_z}_j \psi_j (\mathbf{x})
\end{equation}
where $\zeta_{c_j}$ refers to its SU(3) color state, $q^{s_z}_j$ is its spin-flavor state with polarization $s_z$, and $\psi_j$ is the spatial wave function. For example, the wave function of a red (r) up-quark ($u$) with spin up polarization would be denoted as $\zeta_r  u^\uparrow \psi(\mathbf{x})$.
We now consider the problem routinely faced by event generators in QCD. What is the quantum mechanical probability to create a meson $M$ with arbitrary spin state $j_z$ and momentum $\mathbf{P}_f$ from a particular quark-antiquark pair?
In most applications, the spin state of the quarks is not known, and color states are either unknown or the information is incomplete (e.g.\ because a large-$N_c$ approximation is used). In addition, positions and momenta are often given as if the quarks were classical particles, rather than complete information on their quantum mechanical wave function. We will return to the last point in the next subsection and deal with the color, spin and flavor algebra first.

If quark spin and color states are completely unknown, they have to be treated statistically. Then the coalescence probability of a quark-antiquark pair into a given meson $M$ with given momentum $\mathbf{P}_f$ is
\begin{multline}
   \mathcal{P}_{M,\mathbf{P}_f} = \sum_{J_z} \frac{1}{4}\sum_{s_1,s_2} \\
   \times \frac{1}{N_c^2}\sum_{c_1,c_2} \left| {\Large\langle} \Psi_{M}^{j_z} \Phi_{\mathbf{P}_f}  {\Large|}  \psi^{s_1}_{1,c_1} (\mathbf{x}_1) , \>   \psi^{s_2}_{2,c_2} (\mathbf{x}_2)
    {\Large\rangle}  \right|^2     \, .
\end{multline}
where we average over both quark spin and color degrees of freedom. Note that we will likewise not be interested in the polarization of the meson, and will therefore sum over its total magnetic quantum number $J_z$ as well.  $\Phi_{\mathbf{P}_f}$ represents the plane wave motion of the meson which is dealt with in the next subsection. Note that coordinates for the meson are again suppressed for brevity.

Using the general expressions in Eqs.\ (\ref{eq:mesongeneral}) and (\ref{eq:qwf}) for the wave functions, we can first factor out the probability to form a color singlet
\begin{equation}
    \mathcal{P}_c = \frac{1}{N_c^2}\sum_{c_1,c_2} {\left| \large\langle \zeta_{\mathbf{1}} 
    \large| \zeta_{c_1} \zeta_{c_1} \large\rangle
    \right|^2} = \frac{1}{9} \, .
\end{equation}
It is easy to modify this factor for certain other simple scenarios. For example, if it is known that the two colors form a color singlet, or if it is known that they are prohibited from forming a color singlet (e.g.\ if they emerge from 
a gluon splitting $g\to q\bar q$), the color factor in the probability can be adjusted to $\mathcal{P}_c = 1$ or $0$, respectively.
%In the last line we have factored the probability using the wave function in Eq.\ (\ref{eq:mesongeneral}) in terms of color, flavor, spin and spatial overlap probabilities.
The coalescence probability then becomes
\begin{multline}
   \mathcal{P}_{M,\mathbf{P}_f} = \mathcal{P}_c \mathcal{P}_f \sum_{J_z} \frac{1}{4} \sum_{s_1,s_2} 
   \sum_{\substack{S_z+L_z=J_z \\ S_z'+L_z'=J_z}} 
   \!\!\!\!\!\! C^{J, J_z}_{L,L_z;S,S_z}
   {C^{J, J_z*}_{L,L_z';S,S_z'}}   \\
   \times {\large\langle} \chi_{S}^{S_z} {\large|} s_1, s_2 {\large\rangle}
   {\large\langle}  s_1, s_2 {\large|} \chi_{S}^{S_z'} {\large\rangle}
   \\  \times
   {\Large\langle} Y_{L}^{L_z} H_{M}^{n,L} \Phi_{\mathbf{P}_f} {\Large|}\psi_1(\mathbf{x}_1) \psi_2(\mathbf{x}_2) {\Large\rangle} \\ \times
    {\Large\langle}  \psi_1(\mathbf{x_1}) \psi_2(\mathbf{x}_2) {\Large\rangle} {\Large|}  Y_{L}^{L_z'} H_{M}^{n,L} \Phi_{\mathbf{P}_f} {\Large\rangle} \\
    =  \mathcal{P}_c \mathcal{P}_f \frac{1}{4}\sum_{J_z}  \sum_{S_z+L_z=J_z } \mathcal{P}_{n,L, L_z,\mathbf{P}_f}
    {\left|C^{J, J_z}_{L,L_z;S,S_z}\right|}^2  \, .
\end{multline}
%where we have again suppressed coordinates for the wave functions related to the meson for brevity. 
For the last equality, we have used the fact that we have a complete set of initial spin states $\sum|s_1,s_2|\rangle\langle s_1,s_2|=1$, and we have defined the spatial overlap probability
\begin{equation}
  \mathcal{P}_{n,L, L_z,\mathbf{P}_f} = {\left|  
  {\Large\langle} Y_{L}^{L_z} H^{n,L}_{M} \Phi_{\mathbf{P}_f} {\Large|}\psi_1(\mathbf{x_1}) \psi_2(\mathbf{x}_2) {\Large\rangle}  \right|}^2
\end{equation}
for a meson with quantum numbers $n$, $L$ and $L_z$, and momentum $\mathbf{P}_f$. The flavor overlap probability is $\mathcal{P}_f=1$, except for a few cases of mixing discussed further on.
%since the flavor of the quark and antiquark are known.

The sum of Clebsch-Gordan coefficients can be evaluated to (see App.\ \ref{app:cgs}) 
\begin{equation}
\label{eq:CGsum}
   \sum_{J_z}  {\left|C^{J, J_z}_{L,L_z;S,J_z-L_z}\right|}^2
   = \frac{2J+1}{2L+1} \, .
\end{equation}
Crucially, this result is independent of the value of the magnetic quantum number $L_z$. This fact is key to the often-used rule that only degeneracies, i.e.\ a counting of possible final states, is needed for unpolarized coalescence.
This becomes obvious once we simplify the remaining expression to
\begin{equation}
 \mathcal{P}_{M,\mathbf{P}_f} = \mathcal{P}_c \mathcal{P}_f \frac{1}{4}\frac{2J+1}{2L+1}\sum_{L_z=-L}^L  \mathcal{P}_{n,L, L_z,\mathbf{P}_f}  
 \, .
\end{equation}
The numerator $2J+1$ is precisely the number of spin states for a fixed $J$, while  $4(2L+1)$  is the number of all possible angular momentum states for two spin-$1/2$ particles with given angular momentum $L$.

We finish by introducing the spatial wave function overlap
summed over the magnetic quantum number
\begin{equation}
    \mathcal{P}_{n,L,\mathbf{P}_f} = %\frac{1}{2L+1}
    \sum_{L_z=-L}^L  \mathcal{P}_{n,L, L_z,\mathbf{P}_f}   \, .
\end{equation}
This allows us to arrive at a compact final expression for the recombination probability
\begin{equation}
 \label{eq:Pmf}
 \mathcal{P}_{M,\mathbf{P}_f} = 
 \mathcal{P}_c \mathcal{P}_f
 \frac{2J+1}{4(2L+1)} \mathcal{P}_{n,L,\mathbf{P}_f} 
 \, ,
\end{equation}
which only leaves the spatial wave function overlap to be determined.

\subsection{Spatial Wave Function Overlap}

For a quark and antiquark with masses $m_1$ and $m_2$, and coordinates $\mathbf{x}_1$ and $\mathbf{x}_2$, respectively, we introduce center and relative coordinates 
\begin{equation}
  \mathbf{X} = \frac{m_1}{m_1+m_2} \mathbf{x}_1 + \frac{m_1}{m_1+m_2}\mathbf{x}_2 \, , \quad \Delta \mathbf{x} = \mathbf{x}_1 - \mathbf{x}_2 
  \, ,
\end{equation}
as usual. For momentum variables $\mathbf{k}_i$ we likewise define a total and relative momentum
\begin{equation}
    \mathbf{K} = \mathbf{k}_1+\mathbf{k}_2  \, , \quad  \Delta \mathbf{k} = \frac{m_2}{m_1+m_2}  \mathbf{k}_1 - \frac{m_2}{m_1+m_2} \mathbf{k}_2 
  \, .
\end{equation}
Up to this point, our considerations are independent of the interaction between quarks. However, in order to compute the probabilities $\mathcal{P}_{n,L,\mathbf{P}_f}$, we have to now specify a potential. We choose the harmonic oscillator potential for both its successes and its simplicity. We do not consider spin-spin or spin-orbit interaction terms.

To be precise, we consider the Hamiltonian $\hat H_M = \hat H_0 + \hat H_{q\bar q}$ where
$\hat H_0=\hat K^2/(2M)$ describes the center-of-mass motion, and
\begin{equation}
      \hat H_{q\bar q} = \frac{{(\Delta\hat{k)}}^2} {2 \mu} + \frac{1}{2} \mu \omega^2 {(\Delta\mathbf{x})}^2   \, ,
\end{equation}
where the caret symbol indicates operators. $M=m_1+m_2$ is the total mass, and $\mu = m_1 m_2/(m_1+m_2)$ is the reduced mass of the system. $\omega$ describes the strength of the interaction potential, which will later be determined by experimental data for each meson, as far as possible. In the following, it will turn out to be more convenient to introduce the inverse natural length scale of the oscillator $\nu =\sqrt{\mu \omega/\hbar}$. The value of $\nu$ does depend on the quark flavors under consideration.

Common eigenstates of $\hat H_{q\bar q}$ and the orbital angular momentum operators $\hat {\mathbf L}^2$ and $\hat L_z$ are well known \cite{KORDELLII2022168960}. The complete set of eigenstates is labeled by a radial quantum number $k$ and the usual angular momentum quantum numbers $l$ and $m$. The energy of an eigenstate is given by $E=(N+3/2)\hbar\omega$, where the combined energy quantum number is
\begin{equation}
   N = 2k + l \, .
\end{equation}
We identify $l$ with the orbital angular momentum quantum number $L$ of the meson, and $k$ can be identified with the radial meson quantum number as $n=k+1$. We will use the notation $k$, $l$ in this section.

In Ref.\ \cite{KORDELLII2022168960}, the isotropic 3-D harmonic oscillator problem was solved in  the Wigner phase space approach. As discussed above, the rationale for utilizing this approach is the fact that input for hadronization is often given in a quasi-classical approximation. The coordinates and momenta of quarks, rather than some initial wave functions, are provided. To make sense of this input quantum mechanically, in the absence of additional information, it was assumed in Ref.\ \cite{KORDELLII2022168960} that these phase space coordinates are the most likely values of Gaussian wave packets in phase space. The choice of Gaussian shapes is for simplicity in the absence of additional information. If these phase space values are given by  $(\mathbf{r}_1,\mathbf{p}_1)$, $(\mathbf{r}_2,\mathbf{p}_2)$, respectively,
the Wigner distributions for the quarks are
\begin{equation} 
  \label{eq:qwigner}
  W_{i} \left(\mathbf{x},\mathbf{k}\right) = \frac{1}{\pi^3\hbar^3 } e^{-\frac{1}{2 \delta^2_i}(\mathbf{x}-\mathbf{r_i})^2 - 
   \frac{2 \delta^2_i}{\hbar^2} (\mathbf{k}-\mathbf{p_i})^2 }
  %\label{eq:gaussinput}
\end{equation}
where the widths $\delta_i$, for $i=1,2$, are typically unknown.

The Wigner distributions of angular momentum eigenstates, \emph{averaged} over the magnetic quantum numbers, for the lowest few values of quantum numbers
$k$ and $l$ are \cite{KORDELLII2022168960}
\begin{align}
  W_{00} & =  \frac{1}{\pi^3\hbar^3} e^{-\frac{{(\Delta k)}^2}{\hbar^2 \nu^2}-\nu^2 r{(\Delta x)}^2 }  \\   
  W_{01} & =  W_{00}
     \left( -1 +\frac{2}{3} \nu^2 {(\Delta x)}^2 +\frac{2}{3} \frac{{(\Delta k)}^2}{\hbar^2 \nu^2} \right) \, .
\end{align}
The remaining Wigner distributions used in this work are listed in Appendix \ref{app:wignertable}.
In the Wigner formalism, the probability $\pcal_{klm,\mathbf{P}_f}$ for the coalescence of two quark wave packets into a bound state with quantum numbers $k$, $l$ and magnetic quantum number $m$, and with total momentum $\mathbf{P}_f$ is given by
\begin{multline}
   \label{eq:overlap1}
   \pcal_{klm,\mathbf{P}_f}= 
   {(2\pi\hbar)}^6 \int d^3 \mathbf{x}_1 d^3 \mathbf{x}_2 d^3 \mathbf{k}_1 d^3 \mathbf{k}_2  W_{\mathbf{P}_f} (\mathbf{K})    \\ \times
   W_{klm}  \left( \Delta \mathbf{x} , \Delta \mathbf{k} \right) 
   W_1(\mathbf{x}_1,\mathbf{k}_1) W_2(\mathbf{x}_2,\mathbf{k}_2)   \, .
  %\label{eq:coalp1}
\end{multline}
Here, $W_{\mathbf{P}_f} (\mathbf{K})$ is the phase space distribution for a
plane wave with momentum $\mathbf{P}_f$. As discussed in Ref.\ \cite{KORDELLII2022168960}, the probability to create a meson
in a plane wave state with precise momentum $\mathbf{P}_f$ is 
smeared around $\mathbf{p}_1+\mathbf{p}_2$ by an uncertainty given by the widths of the incoming wave packets. If one so wishes, one could sample the final meson momentum from this distribution. However, since the $\delta_i$ are usually not known, we will take the simple point of view of restoring the semi-classical picture for the total momentum and setting $\mathbf{P}_f=\mathbf{p}_1+\mathbf{p}_2$. This is
formally equivalent to taking $\delta_i \to 0$ for the total momentum distribution.
To this end, we integrate over $\mathbf{P}_f$ in the following to catch the total probability of ending up in a bound state with quantum numbers $k$, $l$ and $m$. This is supplemented by enforcing $\mathbf{P}_f=\mathbf{p}_1+\mathbf{p}_2$. We will also sum over the magnetic quantum number $m$ to define
\begin{equation}
    \label{eq:overlap2}
   \pcal_{kl} = \sum_m \int d^3 \mathbf{P}_f \pcal_{klm,\mathbf{P}_f}  \, 
\end{equation}
as the probability to form a bound state with quantum numbers $k$, $l$.

Combining equations (\ref{eq:qwigner}), (\ref{eq:overlap1}), and (\ref{eq:overlap2}), and performing all integrals except those for relative coordinates and momenta, one can show that 
\begin{multline}
\label{eq:pf}
   \pcal_{kl}=  8 \frac{d_w^3}{d_s^3}   
   \int d^3 \Delta \mathbf{x} \, d^3 \Delta \mathbf{k}  \,
   W_{kl} \left( \Delta \mathbf{x} , \Delta \mathbf{k} \right)   \\ \times
   e^{-\frac{1}{2d_s^2} {\left(\Delta \mathbf{x}- \mathbf{r}\right)}^2} 
   e^{-\frac{2 d_w^2 }{\hbar^2}  {\left(\Delta \mathbf{k}-\mathbf{p}\right)}^2 }    \, .
\end{multline}
Here we have used center and relative coordinates for the average positions and momenta of the wave packets,
\begin{align}
  & & \mathbf{r} &= \mathbf{r}_1 - \mathbf{r}_2 \\
  \mathbf{P}_i &= \mathbf{p}_1+\mathbf{p}_2  \, , 
  &\mathbf{p} &= \frac{m_2}{m_1+m_2} \mathbf{p}_1 - \frac{m_1}{m_1+m_2} \mathbf{p}_2    \, ,
\end{align}
and we have introduced combined and weighted widths
\begin{align}
    d_s &= \sqrt{\delta_1^2 + \delta_2^2} \\
    d_w &= \delta_2 \delta_2 \frac{m_1+m_2}{\sqrt{\delta_1^2 m_1^2 + \delta_2^2 m_2^2}}
\end{align}

We can now return to the expression for the coalescence probability
in Eq.\ (\ref{eq:Pmf}) and state that for quarks, described by Gaussian wave packets, and interacting via a harmonic oscillator potential, the coalescence probability for a given meson can be
written as
\begin{equation}
 %\label{eq:Pmf}
 \mathcal{P}_{M,\mathbf{P}_f} = \delta_{\mathbf{P}_f,\mathbf{P}_i} \,
 \mathcal{P}_c \mathcal{P}_f
 \frac{2J+1}{4} \mathcal{P}_{n-1,l} 
 \, ,
\end{equation}
where the Kronecker-$\delta$ indicates the momentum conservation discussed above, and $\mathcal{P}_{n-1,l}$ is given by Eq.\ (\ref{eq:pf}).
%Note that it is customary to start counting spectroscopic quantum numbers $n$ from 1, while $k$ starts with a value 0.

\subsection{Computing the Probabilities}

In this subsection, we follow the strategy outlined in Ref.\ \cite{KORDELLII2022168960} to compute the $\pcal_{kl}$. The original
work considered equal masses and wave function widths for the two quarks. We present here a generalized version with arbitrary masses and widths.
The probabilities can be conveniently written using generating functions for the 1-D harmonic oscillator. More precisely, we can reduce the probabilities to products of 1-D quasi-probabilities $\hat P_{n'  n}$
\begin{equation}
   \label{eq:pkl}
   \pcal_{kl} =  (2l+1)\!\!\! \sum_{\substack{n_1,n_2,n_3 \\ n'_1,n'_2,n'_3}}  D_{kl}\! \left( \substack{n_1,n_2,n_3 \\ n'_1,n'_2,n'_3} \right)
  % \\  \times  
  \!\!
   \hat P_{n'_1 n_1} \hat P_{n'_2  n_2} \hat P_{n'_3  n_3}\, ,
\end{equation}
where the coefficients $D_{kl}$ are defined in Ref.\ \cite{KORDELLII2022168960}. %and the $\hat P_{n'  n}$ are quasi-probabilities obtained from off-diagonal 1-D Wigner functions.
The $\hat P_{n'  n}$ can be computed from off-diagonal 1-D Wigner functions. They can be written as
\begin{equation}
   \label{eq:wgen2}
   \hat P_{n' \, n} (r,p)= \frac{1}{\sqrt{n! n'!}}\frac{\partial^{n+n'}}{{\partial \alpha^{n'}}{\beta^{n}}} I(\alpha,\beta;r,p)\Bigg|_{\alpha=\beta=0}
   \, 
\end{equation}
with generating functions ($i=1,2,3$)
\begin{multline}
    I(\alpha,\beta;r_i,p_i) =  2 \frac{{d_w}}{d_s}   
   \int d \Delta x \, d \Delta k  \,
   G(\alpha,\beta;\Delta x_i,\Delta k_i)     \\ \times
   e^{-\frac{1}{2d_s^2} {\left(\Delta x_i- r_i \right)}^2} 
   e^{-\frac{2 d_w^2 }{\hbar^2}  {\left(\Delta k_i - p_i \right)}^2 }    \, .
\end{multline}
Note that the previous expression is simply the 1-D versions of Eq.\ (\ref{eq:pf})
with the Wigner distribution replaced by its 1-D generating function
\cite{Curtright:2000ux,Kordell:2022}
\begin{equation}
  \label{eq:wgen}
  G(\alpha,\beta; \Delta x_i, \Delta q_i) = \frac{1}{\pi\hbar} e^{\alpha\beta - \left( \nu \Delta x_i-\frac{\alpha+\beta}{\sqrt{2}}\right)^2 - \left( \frac{\Delta k_i}{\hbar \nu}
  + i\frac{\alpha-\beta}{\sqrt{2}}\right)^2}  \, .
\end{equation}

The integrals over the variables $\Delta x_i$ and $\Delta k_i$ can be readily taken. The generating functions become
\begin{multline}
   \label{eq:I1d}
    I(\alpha,\beta;r_i,p_i) = 2 \frac{\zeta_1}{\zeta_2} \sqrt{\frac{1}{(1+\zeta_1^2)(1+\frac{1}{\zeta_2^2})} }  \\
   \times \exp\left[ \alpha\beta - \frac{\left( \nu r_i - \frac{\alpha+\beta}{\sqrt{2}}\right)^2}{1+\zeta_1} - \frac{\left(\frac{p_i}{\hbar \nu} + i \frac{\alpha-\beta}{\sqrt{2}} \right)^2}{1+\frac{1}{\zeta_2}} \right] \, .
\end{multline}
Note that the coalescence probabilities only depend on the meson size $\nu$ and two dimensionless variables
\begin{equation}
    \zeta_1 = \sqrt{2}\nu d_s \, , \quad \zeta_2 = \sqrt{2}\nu d_w
    \, 
\end{equation}
that encode the effect of the quark wave packet widths $\delta_i$ in relation to $\nu$. Recall that the $\delta_i$ are usually unknown. We are therefore free to make choices that simplify our calculation going forward. To be precise, we choose
$\zeta_1 = \zeta_2 =1 $. The uncertainties arising from this choice have been explored in Ref.\ \cite{Kordell:2022}.
%in Appendix \ref{app:parameters}.

As a result of the particular values chosen for the $\zeta_i$, the generating functions simplify to 
\begin{multline}
    I(\alpha,\beta;r_i,p_i) = e^{-\frac{\nu^2 r_i^2}{2} -\frac{p_i^2}{2\hbar^2 \nu^2}}  \\ \times
      e^{\frac{\alpha}{\sqrt{2}}\left(\nu r_i  + i\frac{p_i}{\hbar \nu} \right)}
     e^{\frac{\beta}{\sqrt{2}}\left(\nu r_i  - i\frac{p_i}{\hbar \nu} \right)}   \, .
\end{multline}
This is precisely the form of the generating function already discussed in Ref.\ \cite{KORDELLII2022168960}. This implies 
that in the case of unequal quark masses we can apply the results from the earlier work for equal masses as long as we (i) use reduced masses, and properly weighted relative and total coordinates and momenta as defined in this work; (ii) assume the relations between wave packet widths implied by the conditions $\zeta_i=1$. We will proceed to simply quote the results from Ref.\ \cite{KORDELLII2022168960} below. If the wave packet widths are known in future applications, and condition (ii) no longer holds, it is straight forward, though more tedious, to proceed with Eq.\ (\ref{eq:I1d}).

To quote our final results, we define the dimensionless distance of the quark-antiquark pair in phase space, and the squared dimensionless angular momentum (note the change in notation compared to \cite{KORDELLII2022168960})
\begin{align}
  s &= \frac{\nu^2 r^2}{2} + \frac{p^2}{2\hbar^2\nu^2}   \\
  t &= \frac{1}{\hbar^2} \left( p^2 r^2 - (\mathbf{p}\cdot \mathbf{r})^2 \right)  \, .
\end{align}
The quasi-probabilities given in Ref.\ \cite{KORDELLII2022168960} are
\begin{multline}
  \hat P_{n' \, n}(r,p) = \frac{e^{-s}}{\sqrt{n! n'!}}  \\ \times \left(\frac{\nu r}{\sqrt{2}} + i \frac{p}{\sqrt{2}\nu} \right)^n
  \left(\frac{\nu r}{\sqrt{2}} - i \frac{p}{\sqrt{2}\nu} \right)^{n'}   \, .
\end{multline}
%The $P_{n' \, n}$ are thus indeed quasi-probabilities and can in general be complex numbers.
We can now then quote the coalescence probabilities into angular momentum eigenstates. Up to $N=2k+l=4$ they read
\cite{KORDELLII2022168960}
\begin{align}
   \pcal_{00} &= e^{-s}   \\
   \pcal_{01} &= e^{-s} s \\
   \pcal_{02} &= \frac{1}{2} e^{-s} \left( \frac{2}{3} s^2 +\frac{1}{3} t\right)  \\
   \pcal_{10} &= \frac{1}{2} e^{-s} \left( \frac{1}{3} s^2 -\frac{1}{3} t\right) \\
   \pcal_{03} &= \frac{1}{3!} e^{-s} \left( \frac{2}{5} s^3 +\frac{3}{5} st\right)  \\
   \pcal_{11} &= \frac{1}{3!} e^{-s} \left( \frac{3}{5} s^3 -\frac{3}{5} st\right) 
   \end{align}
   \begin{align}
    %   \\
    \pcal_{04} &= \frac{1}{4!} e^{-s} \left( \frac{8}{35} s^4 +\frac{24}{35} s^2 t + \frac{3}{35} t^2 \right)  \\
   \pcal_{12} &= \frac{1}{4!} e^{-s} \left( \frac{4}{7} s^4 -\frac{2}{7} s^2t - \frac{2}{7} t^2 \right)  \\
   \pcal_{20} &= \frac{1}{4!} e^{-s} \left( \frac{1}{5} s^4 -\frac{2}{5} s^2 t + \frac{1}{5} t^2 \right ) 
   \, .
\end{align}
This concludes the computation of coalescence probabilities.

%We note that the results echo the 1D-case \cite{Han:2016uhh} in the sense that the total probability to coalesce into a state with energy quantum 
%number $N$ is given by the Poisson probability $e^{-v}v^N/N!$ In addition, our probability distributions distinguish 
%between energy-degenerate angular momentum eigenstates as desired.

\section{Meson States Including Excitations}
\label{sec:3}

\subsection{The Meson Spectrum}

Let us now consider in more detail the meson states we would like to populate.
%The quantum numbers $(k,l,s,j)$ of the bound states (after summation over $m_j$) can be mapped directly onto meson states in the quark model \cite{Zyla:2020zbs}, which are  typically classified in spectroscopic notation $M=n^{2s+1}L_j$, where $L$ denotes orbital angular momentum by the usual 
%letters $S$, $P$, $D$, $F$, etc., for $l=0,1,2,3, \ldots$, resp., and $n=k+1$ is the radial quantum 
%number. $2s+1$ denotes the spin degeneray of the two quark-system. 
We are guided by two principles. First, the set of states should be complete up to a cutoff imposed in quantum numbers, specifically a cutoff on the energy quantum number $N=2k+l$.
Second, coalescing mesons should be physical, i.e.\ experimentally confirmed states, as far as possible.

We start to populate the meson tables with states that are confirmed to match quark model states in the Particle Data Book \cite{ParticleDataGroup:2024cfk}. Note that most
of the data from the Particle Data Group was harvested before the latest version was published. It is rather based on the earlier version Ref.\ \cite{ParticleDataGroup:2020ssz}. It is the latter one, which we will cite widely in this manuscript.

In the light quark sector, these mesons are known, more or less, up to $l=2$ (D-wave) for radial ground states $n=1$, and to some extent for S- and P-waves of the first radially excited level $n=2$. For mesons with charm and bottom quarks, confirmed matching between quark model states and experimentally observed state is much poorer than in the light sector. All confirmed states can be found in black font in the Tables in Appendix \ref{sec:app1}. Tables \ref{tab:lightmeson} and  \ref{tab:lightmesonradial} contain all states utilized for bound states of light quarks. Tables  \ref{tab:cmeson}, \ref{tab:cmesonradial}, and \ref{tab:bmeson},  \ref{tab:bmesonradial} contain mesons with charm quarks and bottom quarks, respectively. 

We proceed to complete these tables by adding quark model states that do not have confirmed experimental counterparts up to $N=2k+l=4$. These mesons are grayed out and marked with a dagger. Before we discuss the properties assumed for these states 
in detail, let us address meson states that are experimentally observed but do not have a confirmed counterpart in the quark model. This includes very light states like the $f_0(500)$ and $a_0(980)$, as well as several heavier states. Various explanations for these extra states have been offered in the literature. Some of them have been interpreted as tetraquarks, hadronic molecules, glueballs, or quark states mixing with glueballs \cite{ParticleDataGroup:2020ssz}. We do not include these states here to avoid possible double counting in the Hilbert space considered.
%Including these states could lead to several issues, including possible double counting 
%in the Hilbert space considered here, a need for production channels other than quark-antiquark recombination, etc. Thus we ignore these states in this study for the sake of unitarity. 
Overall, there are 330 meson states, not counting isospin degrees of freedom, for $N\le 4$. 111 of those are known states, 219 are not experimentally confirmed. On the other hand, there are only 11 experimentally confirmed meson states in the light sector with masses below 2 GeV \cite{ParticleDataGroup:2020ssz}, which do not have quark model counterparts. Excluding these states, even if there are 
misidentifications, should not impact most observables.
%light: 4 per set (no isospin) [9 (w/ isopsin)]
%120 (55 known, 65 unknown) [270] ; k=0 72 (42 known, 30 unknown) [162] , k=1 40 (13 known, 27 unknown) [90] , k=2 8 ( 8 unknown) [18]
%charm: 3 per set (no isospin) [6 w/ isospin]
% 90 (31 known, 59 unknown) [180]  ; k=0 54 (23 known,  31 unknown) [108], k=1 30 (7 known, 23 unknown1) [60]; k=2 6 (1 known, 5 unknown) [12]
%bottom: 4 per set (no isospin) [7 w/ isospin]
%120 (25 known, 95 unknown) [210] ; k=0 72 (16 known, 56 unknown) [126] , k=1 40 (8 known, 32 unknown) [70] , k=2 8 (1 known, 7 unknown) [14]
%Total: 330 states [660 with isospin]: 111 known, 219 unknow

\subsection{Meson Masses}

All mesons, including those not listed in the Particle Data Book, are uniquely identified by their spectroscopic information together 
with their flavor and isospin quantum numbers. Subsequently, unique names and particle Monte Carlo codes for each meson can be assigned following the conventions laid out in the Particle Data Book \cite{ParticleDataGroup:2020ssz}, and they are listed in the tables. A few notable exceptions from the rules (e.g.\ $J=5$ states) are discussed in the captions. To determine the masses of unknown excited states, we use the scaling of quadratic 
mass with both the orbital angular momentum quantum number $L$, and the radial quantum number $n$ \cite{Anisovich:2000kxa}. 
These empirical scaling laws are known to hold to large accuracy and can be motivated in a Regge picture \cite{Chew:1962eu,Blankenbecler:1962ez}. The scaling laws can be written as as
\begin{align}
  M^2 &= M^2_0 +  \alpha k \, ,    \label{eq:reggek} \\
  M^2 &= M^2_0 +  \beta L \, ,   \label{eq:reggel} 
\end{align}
for series of mesons of the same flavor and isospin with varying $k$ and $L$, respectively. We have substituted $k=n-1$ to use a quantum number starting with 0.
The intercept $M_0$ is the mass for the lowest state in a series, and $\alpha$ and $\beta$ are the respective slopes for the $k$-
and $L$-series. It has been suggested that the two slope parameters should be the same for series with the same lowest mass state, 
and that the slope should even be universal for all possible series with different isospin $I$, with a value of 
$\alpha=\beta \approx 1.1$ GeV$^2$ \cite{Forkel:2007cm}. 
However, in this work we keep $\alpha$ and $\beta$ separate and fit them independently for each value of $I$, whenever enough
information is available to do so. We omit the masses of Goldstone bosons from fits.

To determine the unknown masses, we apply the following principles. For unknown $k=0$ states, we determine masses by extrapolation in $L$, keeping $k$ constant. For each flavor and isospin $I$, we fit $M_0$ and $\beta$, as long as at least two masses in the series are known. If the slope can not be determined for some isospin because only one state is known, we revert to the universality argument and the averaged slope from other isospin states with the same $S$, $L$ and $J$ quantum numbers is used. For example, for the slope 
$\beta$ for the series starting with the $f_0$(1710) (a 1$^3$P$_0$ state) the average slopes of the series starting with the 
$a_0$(1450), $K_0^*$(1430) and $f_0$(1370) are used. 
%The numerical closeness of slopes of states of the same angular momentum multiplet with different isospin $I$ is well documented \cite{.}.
%We determine them separately for each series, using the masses of known states, excepting the Goldstone bosons.
%If more than two known states are in a given series we perform a fit. 
%The known masses follow the scaling laws within 20\%, in many cases much more accurately. To be more precise, we use the following rules to determine unknown masses:
We proceed similarly for $k=1$ states. If not enough states are available to fit $L$-series, we use Eq.\ (\ref{eq:reggek}) and the universality
between different isospin series to determine missing $k=1$ states for the lowest values of $L$, until enough information is at hand to fit series in $L$.
Masses of $k=2$, $L=0$ states are established using again Eq.\ (\ref{eq:reggek}), and universality if necessary.
Ultimately, only one intercept mass can not be determined from the rules established above; we have to assume that the mass of the $B_c^*$(2S) is the same
as the known mass of the $B_c$(2S).

\subsection{Meson Decays}

To be phenomenologically useful, we have to specify the decay of each excited meson state into ground state mesons, or more precisely into particles that are typically measured in experiment.  %into into particles thstable under strong and electromagnetic decays. 
For many of the known excited states, the branching ratios for their decays have been measured to some accuracy, and they are documented in the Particle Data Book \cite{ParticleDataGroup:2020ssz}.
However, even some of the confirmed states have poorly measured branching ratios. Their decay channels and branching ratios, as well as decays and branching ratios for all the unconfirmed states, need to be estimated. As a practical matter, we assume that decays will be carried out using PYTHIA 8. This is the case in Hybrid Hadronization, which makes calls to PYTHIA 8 for this task \cite{JETSCAPE:2025wjn}. We will therefore focus on all excited meson states considered here, that are not included in the {\tt ParticleData.xml} file in PYTHIA 8.3 \cite{PYTHIA:github}. 

For unknown decays, we will stick to a few simple principles to estimate available channels and their branching ratios: conservation of quantum numbers, including isospin and $G$-parity, and phase space weights.
To establish possible decay channels the guiding principles are:
\begin{itemize}
    \item Only consider the decays into ensembles of final "stable" particles without specifying possible intermediate states. We only consider pions %\footnote{We keep the $pi^0$ despite its electromagnetic decay as its spectra are often reconstructed statistically; of course, PYTHIA can be instructed to decay $\pi^0$s further}, 
    and the ground state version of kaons, $D$, $D_s$, $B$, $B_s$, $B_c$, $J/\psi$, and $\Upsilon$ mesons as stable. 
    \item Conserve flavor and $G$-parity where applicable. Masses of decaying states must be larger than the sums of masses of decaying particle, but in any case we limit the number of final state particles to 5 for reasons of practicability.
    \item Use the OZI rule \cite{OKUBO1963165,10.1143/PTPS.37.21} to veto decays that create additional strange or heavy quark pairs as valence quarks. 
\end{itemize}
These principles typically lead to allowed decays of light flavor hadrons into either an odd or even number of pions, depending on $G$-parity, and of excited $K$, $D$ and $B$ mesons into ground state $K$, $D$ and $B$ plus pions. Excited $D_s$, $B_s$ and $B_c$ are allowed to decay both into their respective ground states plus pions, as well as into $D+K$, $B+K$, and $B+D$ plus pions, respectively. Likewise, for excited quarkonium and bottomonium states, both the decays into $J/\psi$ plus pions and $\Upsilon$ plus pions, and decays into $DD$ plus pions and $BB$ plus pions, respectively, are considered.

As an example, let us consider the three charge states of the $\rho(1715)$, a $1^3\text{D}_2$ state. Our principles allow the decay channels $\rho(1715) \to \pi\pi\pi$ and $\rho(1715) \to \pi\pi\pi\pi\pi$. 

To estimate branching ratios we first adopt Fermi's idea that the leading effect is given by the phase space volume available for different decay channels \cite{ 10.1143/ptp/5.4.570}. The probability for a decay of a particle with 4-momentum $P$ into $n$ particles of momenta $q_i$ and masses $m_i$ can generally be expressed as
\begin{multline}
    S_n(m_1,\ldots m_n) = \int \prod_{i=1}^n 
    \left( d^4 q_i \delta(q_i^2-m_i^2) \right) \\ \times
    \delta^{(4)}
    \left(P-\sum_i q_i\right) f\left(P;q_1, \ldots q_n\right)  \, ,
\end{multline}
where $f$ is a Lorentz-invariant function of the momenta and masses. Following Srivastava and Sudarshan's implementation of
phase space supremacy \cite{Srivastava:1958ve}. we approximate $f$ by a constant that only depends on the masses of the particles. 
More precisely, the authors argue that for dimensional reasons
\begin{equation}
    f = A \prod_i m_i^2
\end{equation}
where $A$ is an overall constant, which will not be relevant for branching ratios. We then apply the recursion relation discussed in Ref.\ \cite{Srivastava:1958ve} to compute the momentum space integrals up to five particles.
The probability for a decay channel with $n$ decay particles
with masses $m_i$ is then computed by normalization,
\begin{equation}
    p_n(m_1,\ldots m_n) = \frac{S_n(m_1,\ldots m_n) }{
    \sum_{n',m'_j} S_{n'}(m'_1.\ldots m'_{n'})}   \, ,
\end{equation}
where the sum in the denominator runs over all allowed decay channels.

Once the number and flavor of the decay products is fixed. we still need to assign probabilities for different charge states.
For this we consider the isospin of the decay products and project onto the isospin of the decaying excited state.
For $n$ decay particles with isospins $I_1, \ldots, I_n$, the product $\otimes_{i=1}^n (2I_i + 1)$ decouples into several multiplets that contain the decaying particle. Because of our choice of stable particles, the $I_i$ can only represent singlet, doublet and triplet isospin states, and we have restricted ourselves to $n\leq 5$. Likewise, the decaying particle is at most an isospin triplet state. The probabilities can then be worked out using SU(2) algebra, e.g.\ by using Clebsch-Gordan coefficients recursively. If there are several multiplets matching the isospin of a decaying particle, we assume a statistical distribution of probabilities to be in any of these multiplets.

As an example, let us once more consider the decay $\rho(1715) \to \pi\pi\pi$. The $\rho(1715)$ has isospin $I=1$ with three charge states, and the $n=3$ decay products have isospins $I_1=I_2=I_3=1$. Since $3\otimes 3 \otimes 3 = 7 \oplus 5\oplus 5 \oplus 3\oplus 3\oplus 3\oplus 1$, there are three isospin triplets that can match the isospin of the $\rho(1715)$. They can be distinguished, e.g.\ by having two of the pions placed in either a isopspin 2, 1 or 0 state. In absence of further knowledge about decay dynamics we assign $1/3$ probability to each of these options.
To pick the latter case (two pions in an isospin 0 configuration) as a simple example, the three pions couple to make the positive charge, $I_z=+1$ state, through 
\begin{multline}
    {|1,1\rangle}_0 = \sqrt\frac{1}{3} |1,1 ; 1,0 ; 1, 1\rangle   \\ -  \sqrt\frac{1}{3} 
    |1,0 ; 1,0 ; 1,1\rangle +  \sqrt\frac{1}{3} |1,0 ; 1,1 ; 1,-1\rangle  \, ,
\end{multline}
where we use $|I,I_z\rangle$ notation for isospin states.
Similar expressions can be found for the cases that two of the pions are in isospin 2 or 1 configurations, and all three pions make the positive charge state, ${|1,1\rangle}_2$ or ${|1,1\rangle}_1$, respectively. After averaging probabilities over the three possible triplet states ${|1,1\rangle}$, we find $\rho^+(1715) \to \pi^+\pi^+\pi^-$ with probability $p_I=3/5$ and $\rho^+(1715) \to \pi^+\pi^0\pi^0$ with probability $p_I=2/5$. Likewise, we investigate the neutral and negative charge states. The latter mirrors the positive state, and for the neutral state we find 
$\rho^0(1715) \to \pi^+\pi^0\pi^-$ with probability $4/5$ and $\rho^0(1715) \to \pi^0\pi^0\pi^0$ with probability $1/5$.

The branching ratio for a particular decay channel is then given by the product 
%\begin{equation}
    $B = p_n \times p_I$
%\end{equation}
of phase space weight and isospin probability. To clarify once more,
we apply the estimates described here for states for which no experimental information is available. If partial experimental information is available, it takes precedence over our estimates.
All mesons, and their estimated masses, decay channels and branching ratios are documented in a PYTHIA 8-compatible {\tt xml}-file, which is available from the authors upon request.

\section{Implementation and Testing}
\label{sec:4}

\subsection{Recombination in Hybrid Hadronization}

We have implemented the work described here into the Hybrid Hadronization module
in JETSCAPE. We use JETSCAPE 3.6 supplemented with the new {\tt xml}-file with additional meson data. The width $1/\nu_M$  of the harmonic oscillator potential for each ground state meson $M$ and its associated tower of excited states is fixed by the measured charge radius of the ground states, where experimental data is available. We follow the procedure outlined in Ref.\ \cite{Han:2016uhh}. Specifically, using the ground state wave functions for computing averages $\langle\ldots \rangle$, we have
\begin{equation}
    \left\langle r_M^2 \right\rangle = \frac{3}{2\nu_M^2} 
    \frac{Q_1 m_2^2+Q_2 m_1^2}{(m_1+m_2)^2}
\end{equation}
where $Q_i$ is the charge of the quark with mass $m_i$ ($i=1,2$) in meson $M$. 
%Where experimental data is not available, ...
The values for light-light and light-strange pairs are taken from measured charge radii for $pi$ and $K$ reported in Ref.\ \cite{ParticleDataGroup:2024cfk}. Likewise, for open heavy flavor systems we use Ref.\ \cite{Hwang:PRD2010}, and for quarkonia we estimate
utilizing Ref.\ \cite{POVH1990653}.

Quark masses are set to the default values for constituent quarks in PYTHIA 8. Hybrid Hadronization expects a list of partons, with flavor, 4-momentum and 4-position provided, as input. Partons can be accepted with to $\sim$1 GeV virtuality. Gluons are provisionally split into quark-antiquark pairs enforcing local energy-momentum conservation. This splitting is only finalized if one of the decay products is recombining. Otherwise, the gluon is used to build the remnant string system.

The Hybrid Hadronization code goes through all quark-antiquark
pairs in the system that are candidates for recombination, computes their coalescence probabilities, and throws dice to determine if a coalescence happens. Note that color tags are assigned by the shower Monte Carlo and Hybrid Hadronization thus assigns color-sector probabilities $\mathcal{P}_c$ as 1 and 0, respectively for confirmed color singlet and octet configurations, and $1/9$ for all other cases.
Recombined mesons are given energies equal to the sum of energies of the coalescing quark and antiquark, which puts most mesons slightly off the mass-shell. They then decay into stable on-shell particles. Stable particles created directly from coalescence are forced to interact elastically with a nearby hadron with small momentum transfer, putting them on-shell in the final state. This reshuffling of momenta is generally a small effect. It could potentially become uncontrollably large for Goldstone pions and kaons, for which the hadron masses are much smaller than the sum of masses of their constituent quarks. The direct coalescence into ground state pions and kaons is therefore disabled by default in Hybrid Hadronization, but can be enabled by a user. For reasonably large allowed maximum excitation levels $N_\text{max}$ for recombination, the yield of pions and kaons is dominated by decay products. The number $N_\text{max}$ is a parameter controlled by the user.

Flavor mixing is only implemented for a few isospin $I_z=0$ meson for which it has been observed. In those cases, the probabilities $\mathcal{P}_f$ from Sec.\ \ref{sec:2} are adjusted manually. These channels are as follows. For $u\bar u$ and $d\bar d$ going to either of the spin-triplet ground states $\rho$ and $\omega$, we assign $\mathcal{P}_f=0.5$ for both channels. On the other hand, $s\bar s$ pairs in the spin-triplet ground state go exclusively into $\phi$ mesons. For spin singlets, we assign $s\bar s$ to $\eta$ and $\eta'$ with 
$\mathcal{P}_f=2/3$ and $1/3$, respectively. Likewise, 
$u\bar u$ and $d\bar d$ configurations become $\pi^0$, $\eta$ and $\eta'$ with flavor probabilities $1/2$, $1/6$ and $1/3$, respectively.
Other possible mixing, e.g.\ for the $K_{1A}$ and $K_{1B}$ states, is not taken into account here.

The default decay time in PYTHIA 8 is set to $c\tau = 10$ mm. For the current study, decays of all particles considered stable here, are, in addition, switched off manually.

%\subsection{Flavor Mixing}

\subsection{Example: Electron-Positron Annihilation}

\begin{figure}[tb]\centering
  \subfloat{
  \includegraphics[width=\columnwidth]{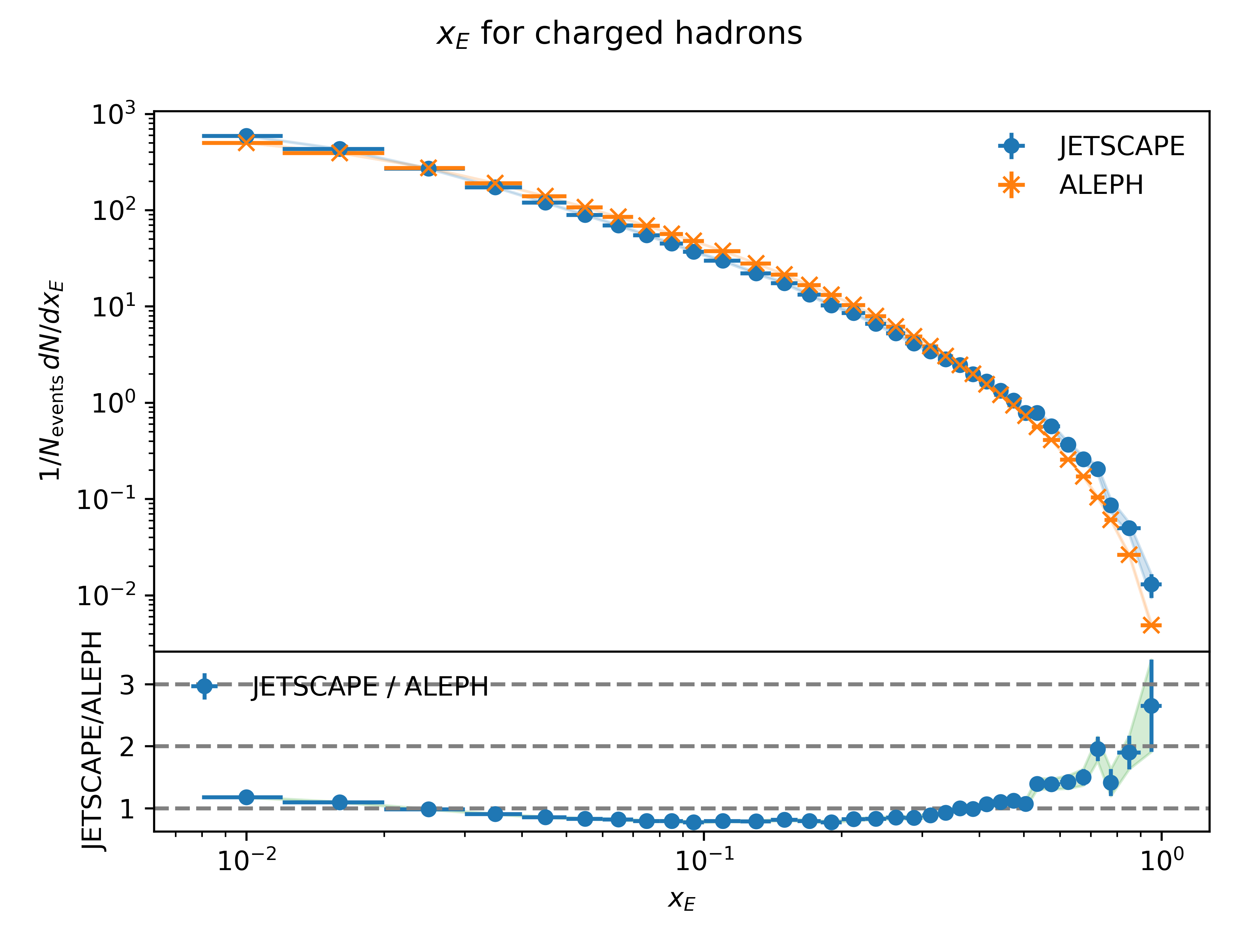}}
  \caption{\label{fig:aleph}
  Spectrum $dN_\text{ch}/dx_E$ of charged hadrons as a function of energy fraction $x_E$ compared to data from the ALEPH collaboration \cite{ALEPH:2003obs}. All available excited meson states ($N_{\text{max}}=4$) are used in recombination and direct Goldstone boson recombination is off. Hadron decays settings and analysis follow the prescription for the ALEPH data set.}
\end{figure}

\begin{figure}[tb]\centering
  \subfloat{
  \includegraphics[width=\columnwidth]{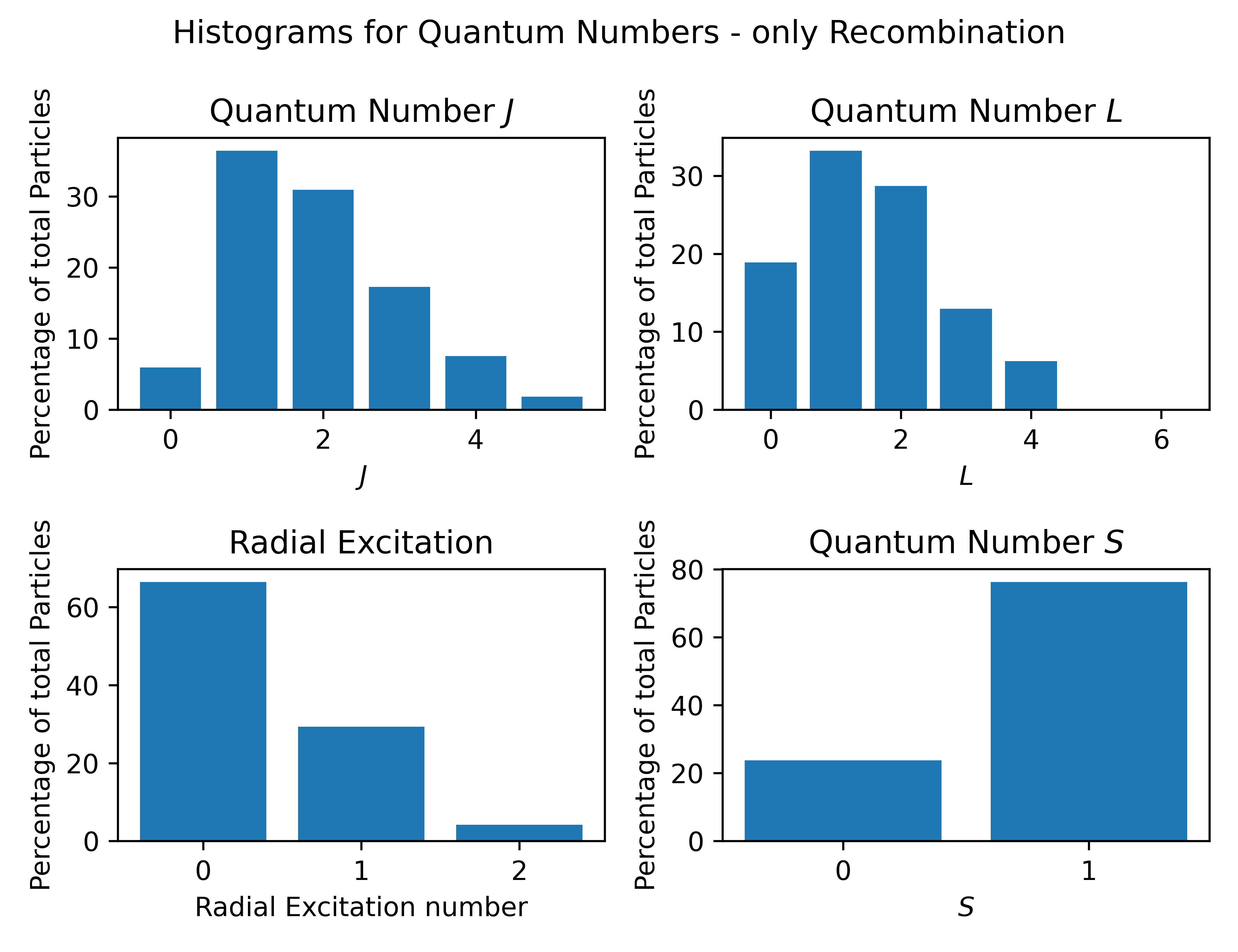}}
  \caption{\label{fig:mult}
  Primordial multiplicities of mesons from recombination only, with full excitation spectrum in Hybrid Hadronization ($N_{\text{max}}=4$) and before decays. Direct Goldstone boson recombination is off.
  The distributions of total meson spin $J$, orbital angular momentum
  $L$, radial excitation number $k$ and spin $S$ are shown.}
\end{figure}

To demonstrate the robustness and some key features of the implementation, we run a set of electron-positron ($e^++e^-$) events at $\sqrt{s}=91.2$ GeV center-of-mass energy. Note that
such a "vacuum-like" system --- compared to, say, a high energy 
nuclear collision --- usually channels only a fraction of its final hadrons through recombination. Most of the hadrons are expected to come from string fragmentation. As the simplest possible system, it is nevertheless an important test. A simple application of Hybrid Hadronization in a more complicated system was carried out in \cite{JETSCAPE:2025wjn}. We
defer more systematic studies of such systems to future work. 

Our setup uses the electron-positron gun in JETSCAPE for the hard process, together with MATTER for parton showers \cite{Majumder:2013re,Cao:2017crw}, and Hybrid Hadronization as described here. MATTER provides a list of partons evolved perturbatively to a low scale $Q_0= 1$ GeV, and it provides expectation values for space-time coordinates for each parton.
Fig.\ \ref{fig:aleph} shows the resulting spectrum of charged hadrons as a function of the energy fraction $x_E=2 E/\sqrt{s}$, the ratio of the energy of a hadron and half the center of mass energy. Both hadrons from recombination and string fragmentation are used for this spectrum to facilitate comparison to experimental data. Data from the ALEPH collaboration taken at the same center-of-mass energy is also shown \cite{ALEPH:2003obs}. The figure demonstrates that the JETSCAPE Monte Carlo setup used here is suitable to roughly describe $e^++e^-$ data. However, precision fits of data are not the main concern here. We simply want to establish that the set of input partons used for Hybrid Hadronization is qualitatively representative of such sets in Monte Carlo simulations.

Next we investigate the quantum numbers of primoridal mesons, i.e., mesons from recombination only, \emph{before} any decays. All mesons, irrespective of charge are recorded as long as they recombine, and no kinematic cuts are applied. Fig.\ \ref{fig:mult} shows the distribution of quantum numbers $J$, $L$, $S$ and $k$  of primordial mesons. We note that P- and D-wave mesons are produced profusely, leading to spin-1 and spin-2 mesons being by far the most common primordial hadrons. On the other hand, radial excitations seem to be disfavored with the first two radially excited states making up only about 30\% of states. Note that the relative abundances are determined by the initial state, i.e.\ by the phase space distribution of partons. In our case, it is the phase space distribution of QCD parton showers as computed by MATTER.

\begin{figure}[tb]\centering
  \subfloat{
  \includegraphics[width=\columnwidth]{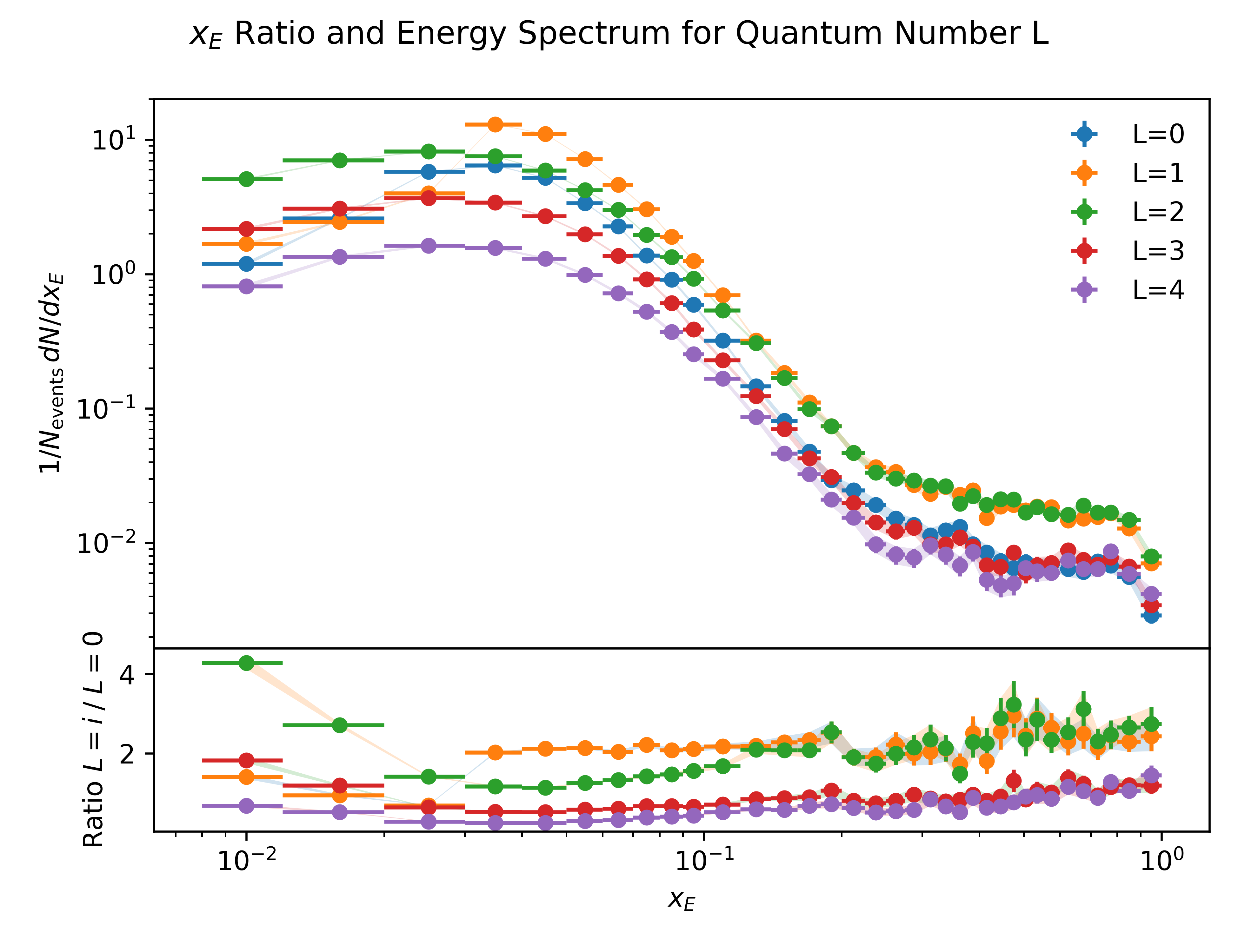}}
  \caption{\label{fig:Lspec}
  Primordial $x_E$-spectrum of mesons from recombination only, with full excitation spectrum in Hybrid Hadronization ($N_{\text{max}}=4$) and before decays. Direct Goldstone boson recombination is off.
  The spectra of all mesons of a particular allowed value of orbital angular momentum $L$ are shown.}
\end{figure}

Fig.\ \ref{fig:Lspec} studies the energy spectrum of mesons from the same calculation as a function of orbital angular momentum $L$ in more detail. %The energy fraction $x_E=2 E/\sqrt{s}$ is the ratio of the energy of a meson and half the center of mass energy.
We find that compared to S-wave mesons, the relative excess of P-wave mesons focuses on intermediate and large $x_E\gtrsim 0.03$,  while for D-wave and larger $L$ states small values of $x_E$ are increasingly favored. The same comparison for the total spin $J$ of the primordial mesons shows a similar but weaker trend and is therefore not shown here.

\begin{figure}[tb]\centering
  \subfloat{
  \includegraphics[width=\columnwidth]{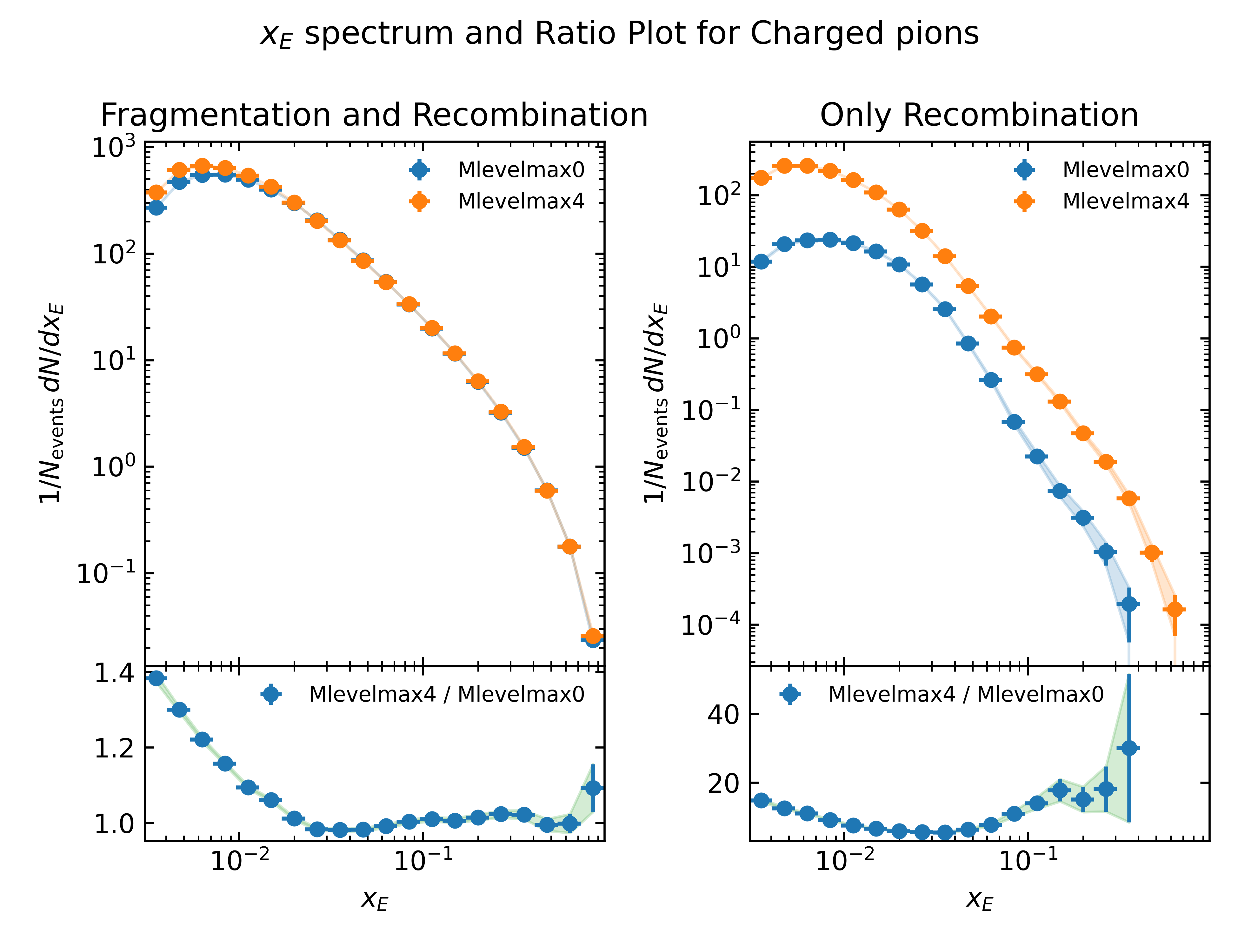}}
  \caption{\label{fig:pions}
  $x_E$-spectrum of charged pions from recombination only (upper right panel) and from all channels (upper left panel) after strong and electromagnetic decays. Spectra with high excitation allowed in recombination ($N_{\text{max}}=4$) and ground states only ($N_{\text{max}}=0$) are compared. The bottom panels show the ratios of spectra $[N_{\text{max}}=4$]/[$N_{\text{max}}=0]$ }
\end{figure}

\begin{figure}[tb]\centering
  \subfloat{
  \includegraphics[width=\columnwidth]{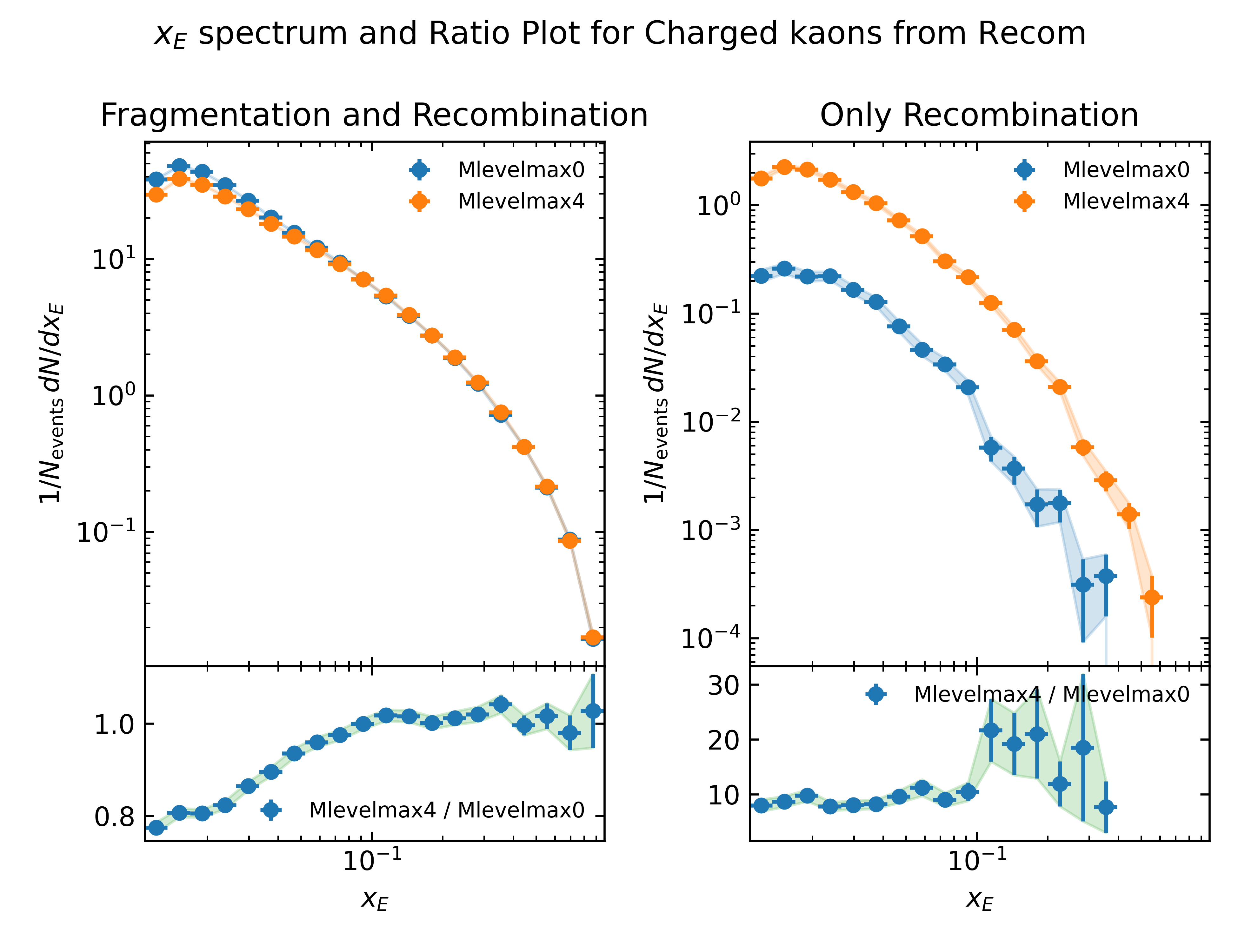}}
  \caption{\label{fig:kaons}
  Same as Fig.\ \ref{fig:pions} for charged kaons.}
\end{figure}

\begin{figure}[tb]\centering
  \subfloat{
  \includegraphics[width=\columnwidth]{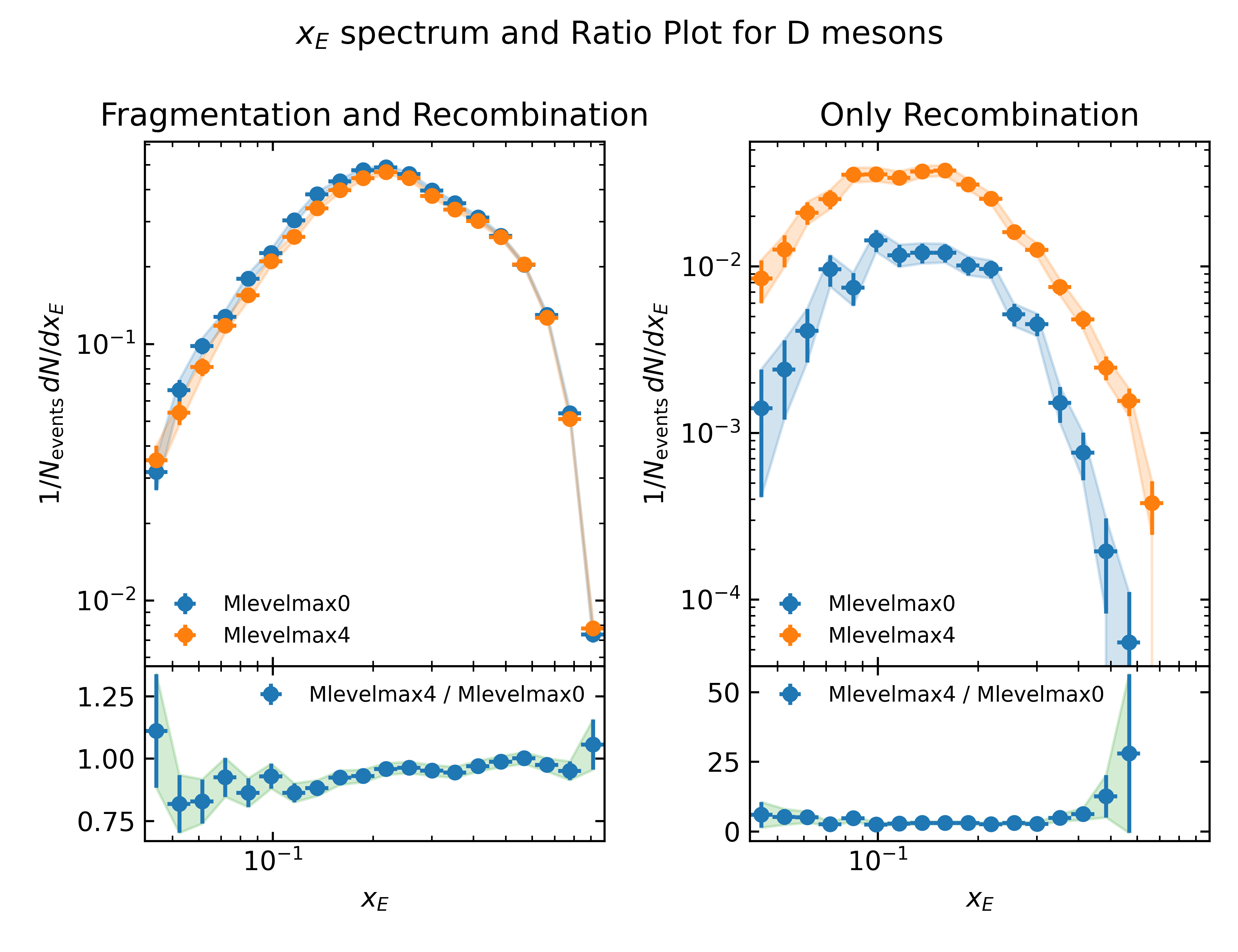}}
  \caption{\label{fig:d}
  Same as Fig.\ \ref{fig:pions} for charged $D$-mesons.}
\end{figure}

\begin{figure}[tb]\centering
  \subfloat{
  \includegraphics[width=\columnwidth]{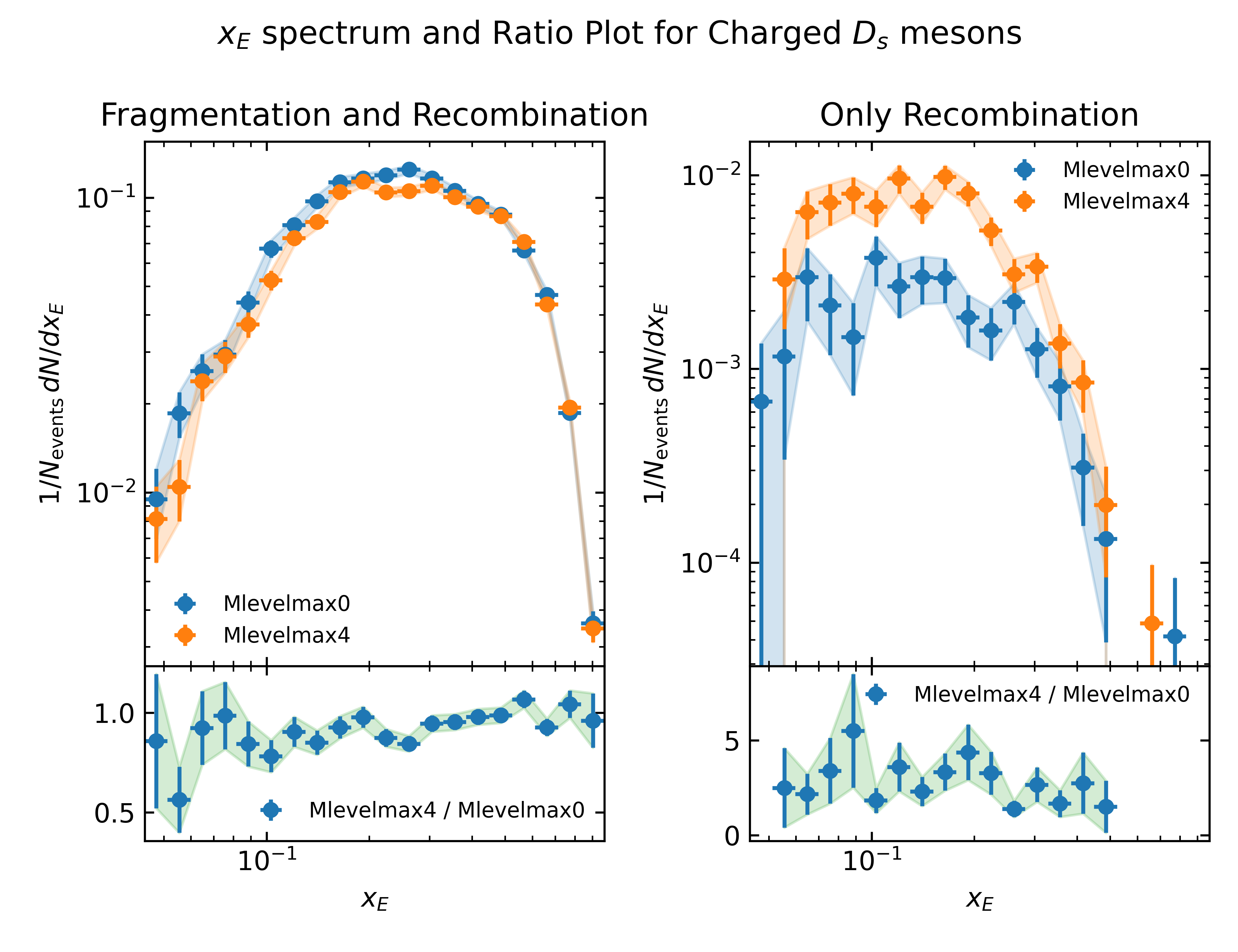}}
  \caption{\label{fig:ds}
  Same as Fig.\ \ref{fig:pions} for $D_s$-mesons.}
\end{figure}

\begin{figure}[tb]\centering
  \subfloat{
  \includegraphics[width=\columnwidth]{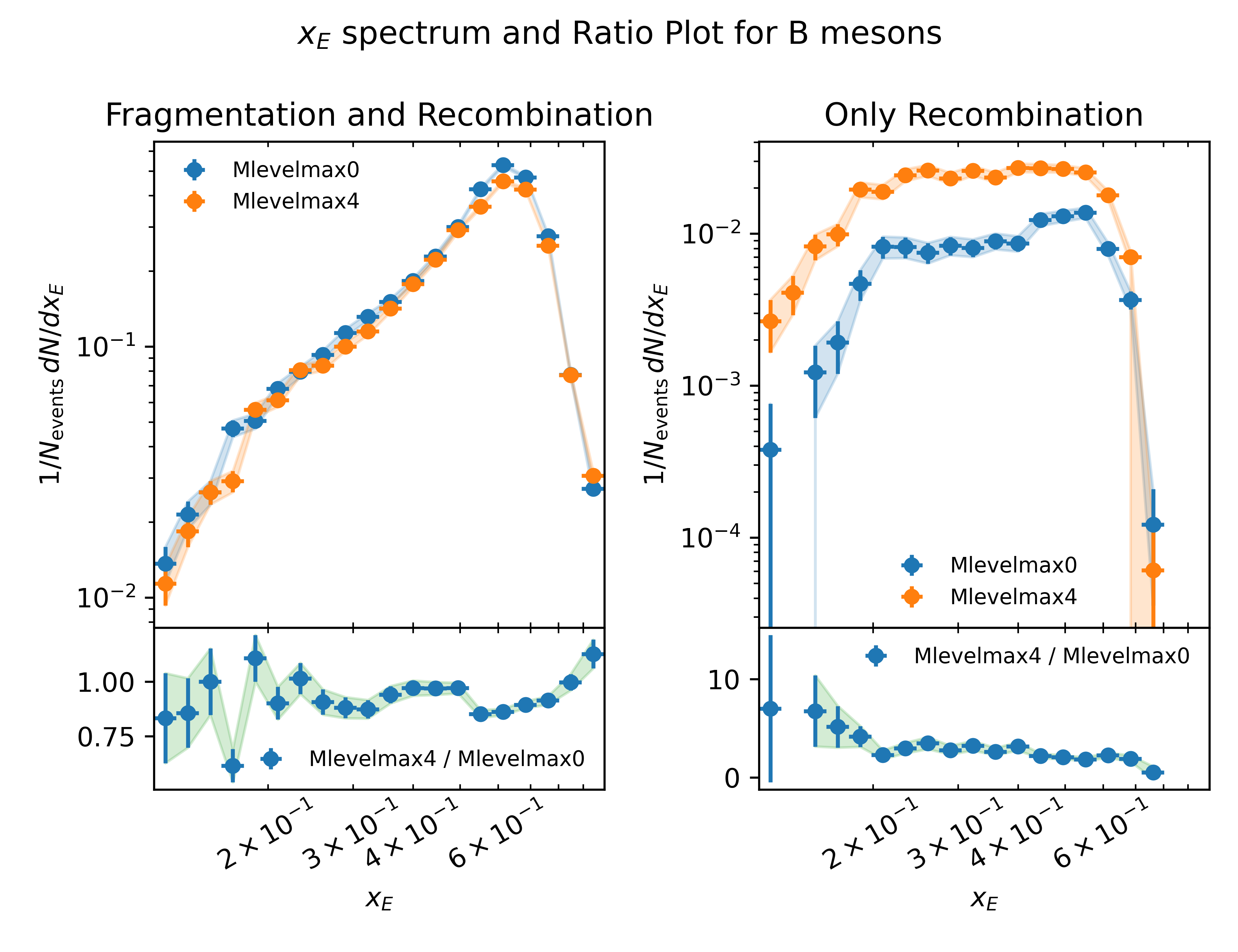}}
  \caption{\label{fig:b}
  Same as Fig.\ \ref{fig:pions} for charged $B$ mesons.}
\end{figure}

In Figs.\ \ref{fig:pions} through \ref{fig:b}, we compare results for recombination with and without highly excited states for a subset of our stable hadrons. For this exercise, decays are switched on. In particular, we show results for charged pions, charged kaons, charged $D$ mesons, $D_s$ mesons, and $B$ mesons, when $N_\text{max}$ for meson recombination is either set to the maximum value of 4 or to the lowest value 0. We show both the result for mesons from recombination only, for which we generally see very large effects with changing $N_\text{max}$. We also show the full spectrum of the pertinent meson including those from string fragmentation. This gives us the total effect, which we expect to be moderate in a system like $e^++e^-$ in which recombination is not the dominant channel.

We find that across all stable mesons here, the yield from recombination increases by a factor 5-10 when going from groundstate only recombination to $N_\text{max}=4$. This increase is of course expected, as the average size of the wave function in phase space increases with each excitation level, increasing the probability for recombination. Therefore, $N_\text{max}$ indeed controls the cutoff between the recombination and string fragmentation channel. The question here is if this change of cutoff affects the total spectrum in a way inconsistent with data. On a practical level, one would hope that in a vacuum-like system the changes are small. This expectation comes from the fact that pure string fragmentation describes many important features in vacuum-like systems like $e^++e^-$. This is indeed confirmed by the panels in Figs.\ \ref{fig:pions} through \ref{fig:b} which show the total stable meson spectra. We see changes of at most 20-25\% for $x_E < 0.1$ and for larger values of $x_E$ spectra appear qualitatively unchanged.

This exercise also demonstrates that the question of how many mesons in a particular system come from either the recombination or fragmentation channel depends on the number of excited states one is including in the former. Basic observables, like momentum or energy spectra in vacuum-like systems, are not very sensitive to this cutoff. What differences remain could most likely be absorbed in a retuning of parameters in the shower Monte Carlo or in string fragmentation. 

As another potentially interesting observation we note that the total yield of pions increases as a function of the $N_\text{max}$ cutoff, while the yield of kaons decreases slightly. I.e., fewer kaons are being produced when string fragmentation is replaced by highly excited states in recombination. The dynamics behind these changes could be complex. They depend on tuning parameters, like the $s\bar s/(u\bar u +d\bar d)$ ratio in string breaking (a parameter in PYTHIA 8, which is we have kept at its default value), they are sensitive to the strength of channels in which strange quarks preferentially form strange baryons, and they are sensitive to highly excited states decaying in systems involving $K\bar K$ pairs, which we have suppressed due to the OZI rule. This calls for future analysis and tuning to data across multiple collision systems, which is beyond the scope of this work.
Such an effort would also require the treatment of baryons on the same footing as mesons.

%String fragmentation ...

\section{Summary and Discussion}

We have discussed a quantum mechanical model for the recombination of quarks and antiquarks into mesons including highly excited states. By implementing this model together with string fragmentation, one arrives at a self-consistent treatment of the hadronization process. Mesons are treated in the non-relativistic quark model with a harmonic oscillator potential, although different potentials could be used if Wigner phase space representations of their eigenstates are known. The phase space formalism allows us to utilize spatial \emph{and} momentum space information of the initial partons in a quantum mechanically consistent fashion. Spin and color information of quarks can be used or treated statistically. The final coalescence probabilities for a given unpolarized meson state only depend on the relative phase space distance
and squared angular momentum of the initial quark-antiquark pair.

We provide a long list of potential excited states up to G-wave states and two radial excitations. Many of these states are not experimentally confirmed but their masses, preferred decay channels and branching ratios are estimated here by utilizing several fundamental principles. The full list of 330 states might be of interest for other applications as well.

We have implemented the improved meson recombination with excited states into the existing Hybrid Hadronization model. We have run electron-positron annihilation events to check the impact of varying the number of allowed excited states. We find that P- and D-wave states play an outsized role in recombination processes. The relative weights of these channels depend on the phase space distribution of the initial partons, which itself is not observable. However, the parton distribution before hadronization used here leads to acceptable descriptions of existing data on hadron spectra.

The highest allowed level of meson excitations, $N_\text{max}$ determines the relative cutoff between recombination and string fragmentation channels. We find that total spectra of identified hadrons in a vacuum-like system like $e^++e^-$
do not vary greatly with this cutoff, making $N_\text{max}$ a great parameter to tune in nucleus-nucleus collisions. We may point out that, in principle, this model does not have any free parameters besides $N_\text{max}$, once the length scales $\nu_M$ are fixed by charge radii data. This suggests that in future work parameters in string fragmentation, or the shower Monte Carlo, could be determined by tuning to data in $e^++e^-$ or $p+p$ collisions. This could be followed up by optimizing $N_\text{max}$ in comparisons to $A+A$ data.

\begin{acknowledgments}
  This work was supported by the U.S.\ National Science Foundation under grants PHY-2111568, PHY-2413003, OAC-2514008 (CSSI:C-SCAPE), as well as OAC-2004571 (CSSI:X-SCAPE) under a subcontract with Wayne State University. C.~M.~K. was supported by
  the U.S.\ Department of Energy under Award No.\ DE-SC0015266.
  W.~L.\ acknowledges support by the Cyclotron Institute.
  REU program funded by NSF award PHY-2447482. Portions of this research were conducted with the advanced computing resources provided by Texas A\&M High Performance Research Computing.
\end{acknowledgments}

\begin{appendix}

%\begingroup
\renewcommand{\arraystretch}{1.25}
\begin{table*}[]
{\centering
\begin{tabular}{|c||c|c|c|c|c|c|c|c|c|c|c|}
\hline
\multicolumn{12}{|c|}{\bf Light/Strange $\mathbf n=1$ ($\mathbf k=0$)}         \\ \hline
\multicolumn{2}{|c|}{}                                                     & \multicolumn{2}{c|}{\textbf{L=0}}                                                                   & \multicolumn{2}{c|}{\textbf{L=1}}                                                                              & \multicolumn{2}{c|}{\textbf{L=2}}                                                                                 & \multicolumn{2}{c|}{\textbf{L=3}}        & \multicolumn{2}{c|}{\textbf{L=4}}                                                              \\ \hline
\multirow{4}{*}{\textbf{\begin{tabular}[c]{@{}c@{}}L=J\\ S=0\end{tabular}}}   & $I=1$           & \multirow{4}{*}{\begin{tabular}[c]{@{}c@{}}xx1\\ $1^1 S_0$\\ $0^{-+}$\end{tabular}} & $\pi$         & \multirow{4}{*}{\begin{tabular}[c]{@{}c@{}}10xx3\\ $1^1 P_1$\\ $1^{+-}$\end{tabular}} & $b_1$(1235)            & \multirow{4}{*}{\begin{tabular}[c]{@{}c@{}}10xx5\\ $1^1 D_2$\\ $2^{-+}$\end{tabular}} & $\pi_2$(1670)             & \multirow{4}{*}{\begin{tabular}[c]{@{}c@{}}10xx7\\ $1^1 F_3$\\ $3^{+-}$\end{tabular}} & {\color{gray}$b_3(2013)^{\dagger}$}     
& \multirow{4}{*}{\begin{tabular}[c]{@{}c@{}} 10xx9 \\ $1^1 G_4$\\ $4^{-+}$\end{tabular}} & {\color{gray}$\pi_{4}(2306)^{\dagger}$} 
      \\ \cline{2-2} \cline{4-4} \cline{6-6} \cline{8-8} \cline{10-10} \cline{12-12}
                                                                              & $I=\frac{1}{2}$ &                                                                                     & $K$           &                                                                                       & {%\color{blue}
                                                                              $K_{1B}$} &                                                                                       & {%\color{blue}
                                                                              $K_2$(1770)} &                                                                                       & {\color{gray}$K_{3B}(2157)^{\dagger}$}   & &   {\color{gray}$K_{4B}(2485)^{\dagger}$}        \\ \cline{2-2} \cline{4-4} \cline{6-6} \cline{8-8} \cline{10-10} \cline{12-12}
                                                                              & $I=0$           &                                                                                     & $\eta$        &                                                                                       & $h_1$(1415)            &                                                                                       & $\eta_2$(1870)            &                                                                                       & {\color{gray}$h_3(2234)^{\dagger}$}       & &  {\color{gray}$\eta_4(2547)^{\dagger}$}      \\ \cline{2-2} \cline{4-4} \cline{6-6} \cline{8-8} \cline{10-10} \cline{12-12}
                                                                              & $I=0$           &                                                                                     & $\eta'$(958)  &                                                                                       & $h_1$(1170)            &                                                                                       & $\eta_2$(1645)            &                                                                                       & {\color{gray}$h_3'(2011)^{\dagger}$}     & &   {\color{gray}$\eta_4'(2320)^{\dagger}$}       \\ \hline
\multirow{4}{*}{\textbf{\begin{tabular}[c]{@{}c@{}}J=L+1\\ S=1\end{tabular}}} & $I=1$           & \multirow{4}{*}{\begin{tabular}[c]{@{}c@{}}xx3\\ $1^3 S_1$\\ $1^{--}$\end{tabular}} & $\rho$(770)   & \multirow{4}{*}{\begin{tabular}[c]{@{}c@{}}xx5\\ $1^3 P_2$\\ $2^{++}$\end{tabular}}   & $a_2$(1320)            & \multirow{4}{*}{\begin{tabular}[c]{@{}c@{}}xx7\\ $1^3 D_3$\\ $3^{--}$\end{tabular}}   & $\rho_3$(1690)            & \multirow{4}{*}{\begin{tabular}[c]{@{}c@{}}xx9\\ $1^3 F_4$\\ $4^{++}$\end{tabular}}   & $a_4$(1970)  
& \multirow{4}{*}{\begin{tabular}[c]{@{}c@{}} xx8$^\ddagger$ \\ $1^3 G_5$\\ $5^{--}$\end{tabular}} & $\rho_5$(2350) 
 \\ \cline{2-2} \cline{4-4} \cline{6-6} \cline{8-8} \cline{10-10} \cline{12-12}
                                                                              & $I=\frac{1}{2}$ &                                                                                     & $K^*$(892)    &                                                                                       & $K^*_2$(1430)          &                                                                                       & $K^*_3$(1780)             &                                                                                       & $K^*_4$(2045) & &  $K_5^*$(2380) \\ \cline{2-2} \cline{4-4} \cline{6-6} \cline{8-8} \cline{10-10} \cline{12-12}
                                                                              & $I=0$           &                                                                                     & $\phi$(1020)  &                                                                                       & $f'_2$(1525)           &                                                                                       & $\phi_3$(1850)            &                                                                                       & $f_4$(2300) & &   {\color{gray}$\phi_{5}(2584)^{\dagger}$}   \\ \cline{2-2} \cline{4-4} \cline{6-6} \cline{8-8} \cline{10-10} \cline{12-12}
                                                                              & $I=0$           &                                                                                     & $\omega$(782) &                                                                                       & $f_2$(1270)            &                                                                                       & $\omega_3$(1670)          &                                                                                       & $f_4$(2050)  & & {\color{gray}$\omega_{5}(2323)^{\dagger}$}    \\ \hline
\multirow{4}{*}{\textbf{\begin{tabular}[c]{@{}c@{}}J=L\\ S=1\end{tabular}}}   & $I=1$           & \multicolumn{2}{c|}{\multirow{8}{*}{}}                                                              & \multirow{4}{*}{\begin{tabular}[c]{@{}c@{}}20xx3\\ $1^3 P_1$\\ $1^{++}$\end{tabular}} & $a_1$(1260)            & \multirow{4}{*}{\begin{tabular}[c]{@{}c@{}}20xx5\\ $1^3 D_2$\\ $2^{--}$\end{tabular}} & {\color{gray}$\rho_2(1715)^{\dagger}$}                         & \multirow{4}{*}{\begin{tabular}[c]{@{}c@{}}20xx7\\ $1^3 F_3$\\ $3^{++}$\end{tabular}} & {\color{gray}$a_3(2072)^{\dagger}$}       
& \multirow{4}{*}{\begin{tabular}[c]{@{}c@{}} 20xx9 \\ $1^3 G_4$\\ $4^{--}$\end{tabular}} & {\color{gray}$\rho_4(2376)^{\dagger}$} 
    \\ \cline{2-2} \cline{6-6} \cline{8-8} \cline{10-10}  \cline{12-12} 
                                                                              & $I=\frac{1}{2}$ & \multicolumn{2}{c|}{}                                                                               &                                                                                       & {%\color{blue}
                                                                              $K_{1A}$} &                                                                                       & {%\color{blue}
                                                                              $K_2$(1820)} &                                                                                       & {\color{gray}$K_{3A}(2160)^{\dagger}$ }   & &   {\color{gray}$K_{4A}(2453)^{\dagger}$}     \\ \cline{2-2} \cline{6-6} \cline{8-8} \cline{10-10} \cline{12-12}
                                                                              & $I=0$           & \multicolumn{2}{c|}{}                                                                               &                                                                                       & $f_1$(1420)            &                                                                                       &  {\color{gray}$\phi_2(1835)^{\dagger}$}                        &                                                                                  
     & {\color{gray}$f_3(2173)^{\dagger}$}         & &    {\color{gray}$\phi_{4}(2464)^{\dagger}$}  \\ \cline{2-2} \cline{6-6} \cline{8-8} \cline{10-10} \cline{12-12}
                                                                              & $I=0$           & \multicolumn{2}{c|}{}                                                                               &                                                                                       & $f_1$(1285)            &                                                                                       & {\color{gray}$\omega_2(1733)^{\dagger}$}                &                                                                                       & 
  {\color{gray}$f_3'(2087)^{\dagger}$}    & &    {\color{gray}$\omega_{4}(2389)^{\dagger}$ }     \\ \cline{1-2} \cline{5-12} 
\multirow{4}{*}{\textbf{\begin{tabular}[c]{@{}c@{}}J=L-1\\ S=1\end{tabular}}} & $I=1$           & \multicolumn{2}{c|}{}                                                                               & \multirow{4}{*}{\begin{tabular}[c]{@{}c@{}}10xx1\\ $1^3 P_0$\\ $0^{++}$\end{tabular}} & $a_0$(1450)            & \multirow{4}{*}{\begin{tabular}[c]{@{}c@{}}30xx3\\ $1^3 D_1$\\ $1^{--}$\end{tabular}} & $\rho$(1700)              & \multirow{4}{*}{\begin{tabular}[c]{@{}c@{}}30xx5\\ $1^3 F_2$\\ $2^{++}$\end{tabular}} & {\color{gray}$a_2(1918)^{\dagger}$}      
& \multirow{4}{*}{\begin{tabular}[c]{@{}c@{}}30xx7 \\ $1^3 G_3$\\ $3^{--}$\end{tabular}} & {\color{gray}$\rho_{3}(2113)^{\dagger}$}   
     \\ \cline{2-2} \cline{6-6} \cline{8-8} \cline{10-10}  \cline{12-12} 
                                                                              & $I=\frac{1}{2}$ & \multicolumn{2}{c|}{}                                                                               &                                                                                       & $K^*_0$(1430)          &                                                                                       & {%\color{blue}
                                                                              $K^*$(1680)} &                                                                                       & {\color{gray}${K_{2}^*}(1897)^{\dagger}$ }    & &  
{\color{gray}${K_{3}^*}(2092)^{\dagger}$}        \\ \cline{2-2} \cline{6-6} \cline{8-8} \cline{10-10} \cline{12-12}
                                                                              & $I=0$           & \multicolumn{2}{c|}{}                                                                               &                                                                                       & $f_0$(1710)            &                                                                                       &  {\color{gray}$\phi_{1}(1931)^{\dagger}$}                       &                                                                                       & 
{\color{gray}$f_2(2129)^{\dagger}$}  & & {\color{gray}$\phi_{3}(2310)^{\dagger}$}     \\ \cline{2-2} \cline{6-6} \cline{8-8} \cline{10-10} \cline{12-12}
                                                                              & $I=0$           & \multicolumn{2}{c|}{}                                                                               &                                                                                       & $f_0$(1370)            &                                                                                       & $\omega$(1650)            &                                                                                       & {\color{gray}$f_2'(1889)^{\dagger}$}    & &    
{\color{gray}$\omega_{3}(2101)^{\dagger}$}        \\ \cline{1-2} \cline{5-12} 
\end{tabular}
}
\caption{\label{tab:lightmeson} Meson states without radial excitation, $n=1$ ($k=0$) in the light and strange quark sector.
%i.e.\ made of only up, down and strange quarks). 
Columns show different orbital angular momentum from 
 $L=0$ to $L=4$. Rows show different quark spin states ($S=0$, $1$) and their possible couplings to the total angular momentum $|L-S|\le J\le L+S$. The three light flavors allow for 5 different total isospin combinations,
$I=1$ (up+down), two $I=1/2$ doublets (up/down+strange) and two different $I=0$ states made from linear combinations of $u\bar u$, $d\bar d$, and $s\bar s$. Only one entry is shown for the $I=1/2$ case as the second doublet consists of antiparticles of the first one.
For each $S$, $L$ and $J$, we show 1) the standard hadron Monte Carlo code for these states as defined in Ref.\ \cite{ParticleDataGroup:2020ssz}, 2) the spectroscopic notation $n^{2S+1}L_J$ for the states, and 3) total spin with parity and C-parity, $J^{PC}$.  The two 'x' placeholders in the Monte Carlo code have to be replaced by the flavor codes for the two quarks in the particular state under consideration, i.e.\ 1 for down quarks, 2 for up quarks and 3 for strange quarks. For more details on these codes, please refer to Ref.\ \cite{ParticleDataGroup:2020ssz}  
$\dagger$: States denoted by $\dagger$ and grayed out should exist in the naive quark model, but have not been unambiguously identified with 
measured states in the particle data book. Their masses and decay channels have been determined as discussed in the text.
%$\dagger$: 
The Monte Carlo codes for $J=5$ states do not follow the standard laid out in Ref.\ \cite{ParticleDataGroup:2020ssz}. The last digit '8' is used to denote $J=5$.
}
\end{table*}
%\endgroup

\section{A Clebsch-Gordon Sum}
\label{app:cgs}

In order to prove Eq.\ (\ref{eq:CGsum}) we express
the Clebsch-Gordon coefficients in-terms of Wigner 3-j symbols \cite{varshalovich1988quantum}
\begin{equation*}
    C^{J, J_z}_{L,L_z;S,S_z} = (-1)^{-(J+J_z+2L)}\sqrt{2J+1} \threej{L}{S}{J}{-L_z}{-S_z}{J_z}   \, .
\end{equation*}
Then one has 
\begin{multline}
    \sum_{J_z} {\left|C^{J, J_z}_{L,L_z;S,S_z}\right|}^2 = \sum_{J_z} (2J+1){\left|\threej{L}{S}{J}{-L_z}{-S_z}{J_z}\right|}^2
    \\=   (2J+1)
    \sum_{J_z} \sum_{S_z}  \delta_{S_z,J_z-L_z}  {\left|\threej{L}{S}{J}{-L_z}{-S_z}{J_z}\right|}^2   = \\   (2J+1)
    \sum_{J_z} \sum_{S_z} {\left|\threej{L}{S}{J}{-L_z}{-S_z}{J_z}\right|}^2 %= \frac{1}{2L+1}
    \, ,
\end{multline}
where the last equality is true because the 3-j symbols are defined as zero if the condition $S_z=J_z-L_z$ is not met.
The well known orthogonality condition for Wigner 3-j symbols \cite{messiah2014quantum}
\begin{equation}
    \sum_{J_z} \sum_{S_z} {\left|\threej{L}{S}{J}{-L_z}{-S_z}{J_z}\right|}^2 = \frac{1}{2L+1}
\end{equation}
proves (\ref{eq:CGsum}).
%So we finally have 
%\begin{equation}
%    \sum_{J_z} {\left|C^{J, J_z}_{L,L_z;S,J_z-L_z}\right|}^2 = \frac{2J+1}{2L+1}
%\end{equation}

\section{3-D Harmonic Oscillator Wigner Functions}
\label{app:wignertable}

The following is a list of the additional Wigner distributions, up to $N=4$, used in this work, except for the cases of quantum numbers 12 and 04 which are very tedious. The Wigner functions are taken from Ref.\ \cite{KORDELLII2022168960}.
\begin{align}
  W_{02} & =  W_{00}
     \left( 1 +  \frac{4}{15} \nu^4 r^4 -\frac{4}{3}\nu^2 r^2 +   \frac{16}{15} \frac{r^2 q^2}{\hbar^2} \right.  \nonumber \\ 
    & \qquad   \left. 
     - \frac{8}{15} \frac{\mathbf{r} \cdot \mathbf{q}}{\hbar}    
     - \frac{4}{3} \frac{q^2}{\hbar^2 \nu^2} +  \frac{4}{15} \frac{q^4}{\hbar^4 \nu^4} \right)\, ,   \\
  W_{10} & = W_{00}
     \left( 1 +  \frac{2}{3} \nu^4 r^4 -\frac{4}{3}\nu^2 r^2
    - \frac{4}{3} \frac{r^2 q^2}{\hbar^2} \right. \nonumber \\
    & \qquad    \left. 
     + \frac{8}{3} \frac{\mathbf{r} \cdot \mathbf{q}}{\hbar}    - \frac{4}{3} \frac{q^2}{\hbar^2 \nu^2} +  \frac{2}{3} \frac{q^4}{\hbar^4 \nu^4} \right)\, ,  \\
  W_{03} &= W_{00}
     \left( -1 +  \frac{8}{105} \nu^6 r^6 -  \frac{4}{5} \nu^4 r^4 + 2 \nu^2 r^2 \right. \nonumber \\
    & \qquad - \frac{16}{5} \frac{r^2 q^2}{\hbar^2}  \left( 1 - \frac{3}{14} \nu^2 r^2  -  \frac{3}{14}\frac{q^2}{\hbar^2 \nu^2} \right) \nonumber \\
     & \qquad  + \frac{8}{5} \frac{\mathbf{r} \cdot \mathbf{q}}{\hbar} \left( 1 -  \frac{2}{7} \nu^2 r^2  
     -  \frac{2}{7}\frac{q^2}{\hbar^2 \nu^2} \right) \nonumber  \\
    & \qquad\qquad    \left.+2 \frac{q^2}{\hbar^2 \nu^2} -  \frac{4}{5} \frac{q^2}{\hbar^4 \nu^4} +  
       \frac{8}{105} \frac{q^6}{\hbar^6 \nu^6}\right)\, ,   %\\
        \end{align}
\begin{align}
  W_{11} &= W_{00}
     \left( -1 +  \frac{4}{15} \nu^6 r^6 -  \frac{22}{15} \nu^4 r^4 + 2 \nu^2 r^2 \right. \nonumber \\
    & \qquad +   \frac{4}{5} \frac{r^2 q^2}{\hbar^2}   \left( 1 -  \frac{1}{3} \nu^2 r^2  -  \frac{1}{3}\frac{q^2}{\hbar^2 \nu^2} \right)\nonumber \\
     & \qquad  - \frac{56}{15} \frac{\mathbf{r} \cdot \mathbf{q}}{\hbar} \left( 1 -  \frac{2}{7} \nu^2 r^2  
     -  \frac{2}{7}\frac{q^2}{\hbar^2 \nu^2} \right) \nonumber  \\
    & \qquad\qquad    \left.+2 \frac{q^2}{\hbar^2 \nu^2} -  \frac{4}{15} \frac{q^4}{\hbar^4 \nu^4} +  
       \frac{22}{15} \frac{q^6}{\hbar^6 \nu^6}\right)  \, ,  \\
  W_{04} &= W_{00}
     \left( 1 + \frac{16}{945} \nu^8 r^8 -\frac{32}{105} \nu^6 r^6 + \frac{8}{5} \nu^4 r^4 \right. \nonumber \\ 
     & \qquad  -\frac{8}{3} \nu^2 r^2
     + \frac{32}{5} \frac{r^2 q^2}{\hbar^2} \left( 1+ \frac{8}{189} \nu^4 r^4 -\frac{3}{7}\nu^2 r^2    \right. \nonumber \\
     & \qquad \left.     -\frac{3}{7}\frac{q^2}{\hbar^2 \nu^2}
     + \frac{8}{189}\frac{q^4}{\hbar^4 \nu^4}
     +\frac{2}{21} \frac{r^2 q^2}{\hbar^2} 
    \right) \nonumber \\
     & \qquad - \frac{16}{5}\frac{{(\mathbf{r}\cdot \mathbf{q})}^2}{\hbar^2}  \left( 1 +\frac{4}{63} \nu^4 r^4 -\frac{4}{7} \nu^2r^2
    -\frac{4}{7}\frac{q^2}{\hbar^2 \nu^2} \right. \nonumber \\
     & \qquad \left. +\frac{4}{63}\frac{q^4}{\hbar^4 \nu^4} +\frac{4}{21} \frac{r^2q^2}{\hbar^2}
     -\frac{2}{63} \frac{{(\mathbf{r}\cdot \mathbf{q})}^2}{\hbar^2}
     \right)  -\frac{8}{3} \frac{q^2}{\hbar^2 \nu^2}  \nonumber \\
     & \qquad 
     \left.  + \frac{8}{5}\frac{q^4}{\hbar^4 \nu^4} -\frac{32}{105} \frac{q^6}{\hbar^6 \nu^6} 
    + \frac{16}{945} \frac{q^8}{\hbar^8\nu^8} 
     \right)  \, ,
\end{align}

\begin{align}
  W_{12} &= W_{00}
     \left(  1  + \frac{8}{105} \nu^8 r^8 - \frac{88}{105} \nu^4 r^6 
     +\frac{38}{15} \nu^4 r^4
    \right. \nonumber \\  
    & \qquad   -\frac{8}{3} \nu^2r^2
    +\frac{12}{15} \frac{r^2 q^2}{\hbar^2} \left(1 +\frac{4}{21}\nu^4 r^4 - \frac{2}{21}\nu^2 r^2
    \right.
    \nonumber \\ & \qquad 
    \left.
     - \frac{2}{21}\frac{q^2}{\hbar^2 \nu^2} +\frac{4}{21}\frac{q^4}{\hbar^4\nu^4} -\frac{4}{7} \frac{r^2 q^2}{\hbar^2}    \right)
    \nonumber \\
    & \qquad
    +\frac{64}{15}\frac{{(\mathbf{r}\cdot\mathbf{q})}^2}{\hbar^2} \left(
    1 -\frac{4}{7}\nu^2 r^2 +\frac{1}{28} \nu^4 r^4
    +\frac{1}{28}\frac{q^4}{\hbar^4\nu^4} \right. \nonumber \\
    &\qquad \left. -\frac{4}{7}\frac{q^2}{\hbar^2\nu^2}
    +\frac{5}{14} \frac{r^2 q^2}{\hbar^2} - \frac{1}{7} \frac{{(\mathbf{r}\cdot\mathbf{q})}^2}{\hbar^2}
    \right)
    -\frac{8}{3}\frac{q^2}{\hbar^2 \nu^2}
     \nonumber \\ & \qquad 
    \left.
    +\frac{38}{15}\frac{q^4}{\hbar^4 \nu^4} 
     -\frac{88}{105} \frac{q^6}{\hbar^6 \nu^6}+ \frac{8}{105}\frac{q^8}{\hbar^8\nu^8}
     \right)
\end{align}

\begin{align}     
  W_{20} &= W_{00}
     \left(  1 +  \frac{4}{15}\nu^4 r^4 -\frac{4}{3} \nu^2 r^2
     +\frac{16}{15} \frac{r^2 q^2}{\hbar^2}  \right.  \\
    & \qquad   \left.
     -\frac{8}{15} \frac{{(\mathbf{r}\cdot\mathbf{q})}^2}{\hbar^2}
    - \frac{4}{3}\frac{q^2}{\hbar^2 \nu^2} +  \frac{4}{15}\frac{q^4}{\hbar^4 \nu^4}
     \right)     
       \, .    %\label{eq:wigfinal2}
\end{align}

%\section{Wave Packet Parameters}
%\label{app:parameters}

\section{Meson Tables}
\label{sec:app1}

%In the following we show the complete list of meson states curated for this project.

%\begingroup
\renewcommand{\arraystretch}{1.25}
\begin{table*}[]
{\centering
\begin{tabular}{|c||c|c|c|c|c|c|c||c|c|}
\hline
\multicolumn{8}{|c||}{\bf Light/Strange $\mathbf n=2$ ($\mathbf k=1$)}  & \multicolumn{2}{c|}{\bf $\mathbf n=3$ ($\mathbf k=2$)}       \\ \hline
\multicolumn{2}{|c|}{}                                                              & \multicolumn{2}{c|}{\textbf{L=0}}                                                                   & \multicolumn{2}{c|}{\textbf{L=1}}                                                                              & \multicolumn{2}{c||}{\textbf{L=2}}                                                                                                                                                 
 & \multicolumn{2}{c|}{\textbf{L=0}}                                                                                                                                                
 \\ \hline
\multirow{4}{*}{\textbf{\begin{tabular}[c]{@{}c@{}}L=J\\ S=0\end{tabular}}}   & $I=1$           & \multirow{4}{*}{\begin{tabular}[c]{@{}c@{}}100xx1\\ $2^1 S_0$\\ $0^{-+}$\end{tabular}} & $\pi$(1300)         & \multirow{4}{*}{\begin{tabular}[c]{@{}c@{}}110xx3\\ $2^1 P_1$\\ $1^{+-}$\end{tabular}} &  {\color{gray}$b_{1}(1668)^{\dagger}$}            & \multirow{4}{*}{\begin{tabular}[c]{@{}c@{}}110xx5\\ $2^1 D_2$\\ $2^{-+}$\end{tabular}} &  {\color{gray}$\pi_{2}(1969)^{\dagger}$}                
& \multirow{4}{*}{\begin{tabular}[c]{@{}c@{}}200xx1\\ $3^1 S_0$\\ $0^{-+}$\end{tabular}}  & {\color{gray}$\pi(1766)^{\dagger}$}  \\ \cline{2-2} \cline{4-4} \cline{6-6} \cline{8-8} \cline{10-10}
                                                                              & $I=\frac{1}{2}$ &                                                                                     & $K$(1460)           &                                                                                       & {\color{gray}$K_{1B}(1796)^{\dagger}$}  &                                                                                       & {\color{gray}$K_{2}(2078)^{\dagger}$}  & & {\color{gray}$K(2042)^{\dagger}$}   \\ \cline{2-2} \cline{4-4} \cline{6-6} \cline{8-8} \cline{10-10}
                                                                               & $I=0$           &                                                                                     & $\eta$(1475)        &                                                                                       &  {\color{gray}$h_{1}(1808)^{\dagger}$}            &                                                                                       & {\color{gray}$\eta_{2}(2089)^{\dagger}$}         & &    {\color{gray}$\eta(1961)^{\dagger}$}           \\ \cline{2-2} \cline{4-4} \cline{6-6} \cline{8-8} \cline{10-10}
                                                                              & $I=0$           &                                                                                     & $\eta'$(1295)  &                                                                                       & {\color{gray}$h_{1}'(1664)^{\dagger}$}             &                                                                                       & {\color{gray}$\eta_{2}'(1966)^{\dagger}$}          & &     {\color{gray}$\eta'(1823)^{\dagger}$}       \\ \hline
\multirow{4}{*}{\textbf{\begin{tabular}[c]{@{}c@{}}J=L+1\\ S=1\end{tabular}}} & $I=1$           & \multirow{4}{*}{\begin{tabular}[c]{@{}c@{}}100xx3\\ $2^3 S_1$\\ $1^{--}$\end{tabular}} & $\rho$(1450)   & \multirow{4}{*}{\begin{tabular}[c]{@{}c@{}}100xx5\\ $2^3 P_2$\\ $2^{++}$\end{tabular}}   & $a_2$(1700)            & \multirow{4}{*}{\begin{tabular}[c]{@{}c@{}}100xx7\\ $2^3 D_3$\\ $3^{--}$\end{tabular}}   & {\color{gray}$\rho_{3}(1918)^{\dagger}$}       & \multirow{4}{*}{\begin{tabular}[c]{@{}c@{}}200xx3\\ $3^3 S_1$\\ $1^{--}$\end{tabular}} &  {\color{gray}$\rho(1901)^{\dagger}$}     \\ \cline{2-2} \cline{4-4} \cline{6-6} \cline{8-8} \cline{10-10}
                                                                              & $I=\frac{1}{2}$ &                                                                                     & $K^*$(1410)    &                                                                                       & $K^*_2$(1980)          &                                                                                       & {\color{gray}${K_{3}^*}(2419)^{\dagger}$}      & &   {\color{gray}${K^*}(1783)^{\dagger}$}       \\ \cline{2-2} \cline{4-4} \cline{6-6} \cline{8-8} \cline{10-10}
                                                                              & $I=0$           &                                                                                     & $\phi$(1680)  &                                                                                       & $f_2$(1950)           &                                                                                       & {\color{gray}$\phi_{3}(2187)^{\dagger}$}     & &    {\color{gray}$\phi(2146)^{\dagger}$}     \\ \cline{2-2} \cline{4-4} \cline{6-6} \cline{8-8}  \cline{10-10}
                                                                              & $I=0$           &                                                                                     & $\omega$(1420) &                                                                                       & $f_2$(1640)            &                                                                                       & {\color{gray}$\omega_{3}(1834)^{\dagger}$}     & &   {\color{gray}$\omega(1850)^{\dagger}$}      \\ \hline
\multirow{4}{*}{\textbf{\begin{tabular}[c]{@{}c@{}}J=L\\ S=1\end{tabular}}}   & $I=1$           & \multicolumn{2}{c|}{\multirow{8}{*}{}}                                                              & \multirow{4}{*}{\begin{tabular}[c]{@{}c@{}}120xx3\\ $2^3 P_1$\\ $1^{++}$\end{tabular}} & $a_1$(1640)            & \multirow{4}{*}{\begin{tabular}[c]{@{}c@{}}120xx5\\ $2^3 D_2$\\ $2^{--}$\end{tabular}} & {\color{gray}${\rho_{2}}(1945)^{\dagger}$}               &  \multicolumn{2}{c}{}              \\ \cline{2-2} \cline{6-6} \cline{8-8} 
                                                                              & $I=\frac{1}{2}$ & \multicolumn{2}{c|}{}                                                                               &                                                                                       &  {\color{gray}${K_{1A}}(1750)^{\dagger}$} &                                                                                       & {\color{gray}${K_{2}}(2038)^{\dagger}$}   & \multicolumn{2}{c}{}       \\ \cline{2-2} \cline{6-6} \cline{8-8} 
                                                                              & $I=0$           & \multicolumn{2}{c|}{}                                                                               &                                                                                       &  {\color{gray}${f_{1}}(1766)^{\dagger}$}            &                                                                                       & {\color{gray}${\phi_{2}}(2052)^{\dagger}$}           & \multicolumn{2}{c}{}               \\ \cline{2-2} \cline{6-6} \cline{8-8} 
                                                                              & $I=0$           & \multicolumn{2}{c|}{}                                                                               &                                                                                       &  {\color{gray}${f_{1}'}(1659)^{\dagger}$}           &                                                                                       &  {\color{gray}${\omega_{2}}(1961)^{\dagger}$}                     & \multicolumn{2}{c}{}    \\ \cline{1-2} \cline{5-8} 
\multirow{4}{*}{\textbf{\begin{tabular}[c]{@{}c@{}}J=L-1\\ S=1\end{tabular}}} & $I=1$           & \multicolumn{2}{c|}{}                                                                               & \multirow{4}{*}{\begin{tabular}[c]{@{}c@{}}110xx1\\ $2^3 P_0$\\ $0^{++}$\end{tabular}} &  {\color{gray}${a_{0}}(1854)^{\dagger}$}          & \multirow{4}{*}{\begin{tabular}[c]{@{}c@{}}130xx3\\ $2^3 D_1$\\ $1^{--}$\end{tabular}} &  {\color{gray}${\rho_{1}}(2129)^{\dagger}$}          & \multicolumn{2}{c}{}          \\ 
\cline{2-2} \cline{6-6} \cline{8-8}  
                                                                              & $I=\frac{1}{2}$ &         \multicolumn{2}{c|}{}  &   &  {\color{gray}${K_{0}^*}(1839)^{\dagger}$}         &                                                                                       &  {\color{gray}${K_{1}^*}(2115)^{\dagger}$} &  \multicolumn{2}{c}{}   \\ \cline{2-2} \cline{6-6} \cline{8-8} 
                                                                              &  $I=0$           &         \multicolumn{2}{c|}{}       &  &  {\color{gray}${f_{0}}(2064)^{\dagger}$}           &                                                                                       &  {\color{gray}${\phi_{1}}(2314)^{\dagger}$}           &   \multicolumn{2}{c}{}            \\ \cline{2-2} \cline{6-6} \cline{8-8} 
                                                                              & $I=0$           &          \multicolumn{2}{c|}{}                                                                         &                                                                                       &  {\color{gray}${f_{0}'}(1792)^{\dagger}$}            &                                                                                       &  {\color{gray}${\omega_{1}}(2075)^{\dagger}$}      & \multicolumn{2}{c}{}                       \\ \cline{1-2} \cline{5-8} 
\end{tabular}
}
\caption{\label{tab:lightmesonradial} Mesons in the first radially excited state $n=2$ ($k=1$) in the light quark sector, up to $L=2$, and in the second radially excited state $n=3$ ($k=2$) for $L=0$. 
The table is organized similar to Tab.\ \ref{tab:lightmeson}.
}
\end{table*}
%\endgroup

In the following tables, we summarize the excited meson states we curated for this project, with up to $N=2k+l=4$. Grayed out states with a dagger symbol are not in the particle data book. In that case, masses quoted in parentheses are the masses extrapolated by us. Tabs.\ \ref{tab:lightmeson} and \ref{tab:lightmesonradial} contain mesons with only light and strange quarks, with and without
radial excitations, respectively. Tables \ref{tab:cmeson} and \ref{tab:cmesonradial} contain the same information for mesons with one or two charm quarks, and Tabs.\ \ref{tab:bmeson} and \ref{tab:bmesonradial}, respectively, list all mesons with one or two bottom quarks.

%\begingroup
\renewcommand{\arraystretch}{1.25}
\begin{table*}[]
{\centering
\begin{tabular}{|c||c|c|c|c|c|c|c|c|c|c|c|}
\hline
\multicolumn{12}{|c|}{\bf Charm $\mathbf n=1$ ($\mathbf k=0$)}         \\ \hline
                                \multicolumn{2}{|c|}{}                             & \multicolumn{2}{c|}{\textbf{L=0}}                                                                  & \multicolumn{2}{c|}{\textbf{L=1}}                                                                                              & \multicolumn{2}{c|}{\textbf{L=2}}                                                                                           & \multicolumn{2}{c|}{\textbf{L=3}}        & \multicolumn{2}{c|}{\textbf{L=4}}                                                    \\ \hline
\multirow{3}{*}{\textbf{\begin{tabular}[c]{@{}c@{}}L=J\\ S=0\end{tabular}}}   & $I=0$           & \multirow{3}{*}{\begin{tabular}[c]{@{}c@{}}xx1\\ $1^1 S_0$\\ $0^{-+}$\end{tabular}} & $\eta_c(1S)$ & \multirow{3}{*}{\begin{tabular}[c]{@{}c@{}}10xx3\\ $1^1 P_1$\\ $1^{+-}$\end{tabular}} & $h_c(1P)$                              & \multirow{3}{*}{\begin{tabular}[c]{@{}c@{}}10xx5\\ $1^1 D_2$\\ $2^{-+}$\end{tabular}} & {\color{gray}$\eta_c(1D)(3993)^{\dagger}$}                              & \multirow{3}{*}{\begin{tabular}[c]{@{}c@{}}10xx7\\ $1^1 F_3$\\ $3^{+-}$\end{tabular}} & 
  {\color{gray}$h_{c}(1F)(4412)^{\dagger}$} & \multirow{3}{*}{\begin{tabular}[c]{@{}c@{}}10xx9\\ $1^1 G_3$\\ $4^{-+}$\end{tabular}} &  {\color{gray}$\eta_c(1G)(4795)^{\dagger}$} \\  \cline{2-2} \cline{4-4} \cline{6-6} \cline{8-8} \cline{10-10} \cline{12-12}    
                                                                              & $I=\frac{1}{2}$ &                                                                                     & $D$          &                                                                                       & $D_1$(2420)                            &                                                                                    &                  {\color{gray}$D_2(2867)^{\dagger}$}                    &                                                                                       & {\color{gray}$D_3(3253)^{\dagger}$} & & {\color{gray}$D_4(3598)^{\dagger}$}\\ \cline{2-2} \cline{4-4} \cline{6-6} \cline{8-8} \cline{10-10} \cline{12-12} 
                                                                              & $I=0$           &                                                                                     & $D_s^\pm$    &                                                                                       & $D_{s1}^\pm$(2536)                     &                                                                                       &  {\color{gray}$D^\pm_{s2}(2998)^{\dagger}$}                                    &                                                                                       & {\color{gray}$D^\pm_{s3}(3398)^{\dagger}$} & & {\color{gray}$D^\pm_{s4}(3756)^{\dagger}$} \\ \hline
\multirow{3}{*}{\textbf{\begin{tabular}[c]{@{}c@{}}J=L+1\\ S=1\end{tabular}}} & $I=0$           & \multirow{3}{*}{\begin{tabular}[c]{@{}c@{}}xx3\\ $1^3 S_1$\\ $1^{--}$\end{tabular}} & $J/\Psi(1S)$ & \multirow{3}{*}{\begin{tabular}[c]{@{}c@{}}xx5\\ $1^3 P_2$\\ $2^{++}$\end{tabular}}   & $\chi_{c2}(1P)$                        & \multirow{3}{*}{\begin{tabular}[c]{@{}c@{}}xx7\\ $1^3 D_3$\\ $3^{--}$\end{tabular}}   & {\color{gray}$\psi_{3}(1D)(3963)^{\dagger}$}                                 & \multirow{3}{*}{\begin{tabular}[c]{@{}c@{}}xx9\\ $1^3 F_4$\\ $4^{++}$\end{tabular}}   &  
 {\color{gray}$\chi_{c4}(1F)(4332)^{\dagger}$} &  \multirow{3}{*}{\begin{tabular}[c]{@{}c@{}} xx8$^\ddagger$ \\ $1^3 G_5$\\ $5^{--}$\end{tabular}}  &  {\color{gray}$\psi_{5}(1G)(4672)^{\dagger}$}   \\
\cline{2-2} \cline{4-4} \cline{6-6} \cline{8-8} \cline{10-10} \cline{12-12} 
                                                                              & $I=\frac{1}{2}$ &                                                                                     & $D^*$        &                                                                                       & $D_2^*$(2460)                          &                                                                                       & $D_3^*$(2750)                       &                                                                                       & 
{\color{gray}$D^*_{4}(3053)^{\dagger}$} & & {\color{gray}$D^*_{5}(3329)^{\dagger}$} \\ \cline{2-2} \cline{4-4} \cline{6-6} \cline{8-8} \cline{10-10} \cline{12-12} 
                                                                              & $I=0$           &                                                                                     & $D_s^{*\pm}$ &                                                                                       & $D_{s2}^{*\pm}$(2573)                  &                                                                                       & $D_{s3}^{*\pm}$(2860)               &                                                                                       & {\color{gray}$D^{*\pm}_{s4}(3168)^{\dagger}$} & & {\color{gray}$D^{*\pm}_{s5}(3449)^{\dagger}$}\\ \hline
\multirow{3}{*}{\textbf{\begin{tabular}[c]{@{}c@{}}J=L\\ S=1\end{tabular}}}   & $I=0$           & \multicolumn{2}{c|}{\multirow{6}{*}{}}                                                             & \multirow{3}{*}{\begin{tabular}[c]{@{}c@{}}20xx3\\ $1^3 P_1$\\ $1^{++}$\end{tabular}} & $\chi_{c1}(1P)$                        & \multirow{3}{*}{\begin{tabular}[c]{@{}c@{}}20xx5\\ $1^3 D_2$\\ $2^{--}$\end{tabular}} & $\psi_2$(3823)                      & \multirow{3}{*}{\begin{tabular}[c]{@{}c@{}}20xx7\\ $1^3 F_3$\\ $3^{++}$\end{tabular}} & {\color{gray}$\chi_{c3}(1F)(4112)^{\dagger}$} & 
\multirow{3}{*}{\begin{tabular}[c]{@{}c@{}} 20xx9 \\ $1^3 G_4$\\ $4^{--}$\end{tabular}} & {\color{gray}$\psi_{4}(1G)(4382)^{\dagger}$} \\ \cline{2-2} \cline{6-6} \cline{8-8} \cline{10-10} \cline{12-12} 
                                                                              & $I=\frac{1}{2}$ & \multicolumn{2}{c|}{}                                                                              &                                                                                       & $D_1$(2430)                            &                                                                                       & {\color{gray}$D_{2}(2864)^{\dagger}$}                                 &                                                                                       & {\color{gray}$D_{3}(3240)^{\dagger}$} & &  {\color{gray}$D_{4}(3576)^{\dagger}$} \\ \cline{2-2} \cline{6-6} \cline{8-8} \cline{10-10} \cline{12-12} 
                                                                              & $I=0$           & \multicolumn{2}{c|}{}                                                                              &                                                                                       & {%\color{magenta}
                                                                              $D_{s1}^\pm$(2460)}    &                                                                                       &  {\color{gray}$D^\pm_{s2}(2889)^{\dagger}$}                                  &                                                                                       &  {\color{gray}$D^\pm_{s3}(3262)^{\dagger}$}  & & {\color{gray}$D^\pm_{s4}(3597)^{\dagger}$}  \\ \cline{1-2} \cline{5-12} 
\multirow{3}{*}{\textbf{\begin{tabular}[c]{@{}c@{}}J=L-1\\ S=1\end{tabular}}} & $I=0$           & \multicolumn{2}{c|}{}                                                                              & \multirow{3}{*}{\begin{tabular}[c]{@{}c@{}}10xx1\\ $1^3 P_0$\\ $0^{++}$\end{tabular}} & $\chi_{c0}(1P)$                        & \multirow{3}{*}{\begin{tabular}[c]{@{}c@{}}30xx3\\ $1^3 D_1$\\ $1^{--}$\end{tabular}} & $\psi$(3770)                        & \multirow{3}{*}{\begin{tabular}[c]{@{}c@{}}30xx5\\ $1^3 F_2$\\ $2^{++}$\end{tabular}} & {\color{gray}$\chi_{c2}(1F)(4095)^{\dagger}$} & \multirow{3}{*}{\begin{tabular}[c]{@{}c@{}}30xx7 \\ $1^3 G_3$\\ $3^{--}$\end{tabular}} & {\color{gray}$\psi_{3}(1G)(4396)^{\dagger}$}\\ \cline{2-2} \cline{6-6} \cline{8-8} \cline{10-10} \cline{12-12}
                                                                              & $I=\frac{1}{2}$ & \multicolumn{2}{c|}{}                                                                              &                                                                                       & $D_0^*$(2300)                          &                                                                                       & {\color{gray}$D^*_{1}(2824)^{\dagger}$}                                  &                                                                                       & {\color{gray}$D^*_{2}(3265)^{\dagger}$} & &  {\color{gray}$D^*_{3}(3653)^{\dagger}$} \\ \cline{2-2} \cline{6-6} \cline{8-8} \cline{10-10} \cline{12-12} 
                                                                              & $I=0$           & \multicolumn{2}{c|}{}                                                                              &                                                                                       & {%\color{magenta}
                                                                              $D_{s0}^{*\pm}$(2317)} &                                                                                       & {%\color{blue}
                                                                              $D_{s1}^{*\pm}$(2860)} &                                                                                       & {\color{gray}$D^{*\pm}_{s2}(3315)^{\dagger}$} & & {\color{gray}$D^{*\pm}_{s3}(3715)^{\dagger}$} \\ \cline{1-2} \cline{5-12} 
\end{tabular}
}
\caption{\label{tab:cmeson} Same as Tab.\ \ref{tab:lightmeson} for bound states of charm quarks with light quarks (up, down, strange) and charmonium states. Two isospin $I=1/2$ (charm+up/down) and two different $I=0$ states ($c\bar c$ and charm+strange) are possible in the quark model.}
\end{table*}
%\endgroup

%\begingroup
\renewcommand{\arraystretch}{1.25}
\begin{table*}[]
{\centering
\begin{tabular}{|c||c|c|c|c|c|c|c||c|c|}
\hline
\multicolumn{8}{|c||}{\bf Charm $\mathbf n=2$ ($\mathbf k=1$)}  & \multicolumn{2}{c|}{\bf $\mathbf n=3$ ($\mathbf k=2$)}       \\ \hline
\multicolumn{2}{|c|}{}                                                              & \multicolumn{2}{c|}{\textbf{L=0}}                                                                   & \multicolumn{2}{c|}{\textbf{L=1}}                                                                              & \multicolumn{2}{c||}{\textbf{L=2}}                                                                                                                                                 
 & \multicolumn{2}{c|}{\textbf{L=0}}                                                                                                                                                
 \\ \hline
\multirow{3}{*}{\textbf{\begin{tabular}[c]{@{}c@{}}L=J\\ S=0\end{tabular}}}   & $I=0$           & \multirow{3}{*}{\begin{tabular}[c]{@{}c@{}}100xx1\\ $\mathbf{2^1 S_0}$\\ $0^{-+}$\end{tabular}} & $\eta_c$(2S)         & \multirow{3}{*}{\begin{tabular}[c]{@{}c@{}}110xx3\\ $\mathbf{2^1 P_1}$\\ $1^{+-}$\end{tabular}} &  {\color{gray}$h_{c}(2P)(3919)^{\dagger}$}            & \multirow{3}{*}{\begin{tabular}[c]{@{}c@{}}110xx5\\ $\mathbf{2^1 D_2}$\\ $2^{-+}$\end{tabular}} &  {\color{gray}$\eta_{c}(2D)(4182)^{\dagger}$}                
& \multirow{3}{*}{\begin{tabular}[c]{@{}c@{}}200xx1\\ $\mathbf{3^1 S_0}$\\ $0^{-+}$\end{tabular}}  & {\color{gray}$\eta_c(3S)(4189)^{\dagger}$}  \\ \cline{2-2} \cline{4-4} \cline{6-6} \cline{8-8} \cline{10-10}
                                                                              & $I=\frac{1}{2}$ &                                                                                     & {\color{gray}$D(2796)^{\dagger}$}           &                                                                                       &  {\color{gray}$D_1(3154)^{\dagger}$}    &                                                                                       & {\color{gray}$D_{2}(3476)^{\dagger}$}  & & {\color{gray}$D(3484)^{\dagger}$}   \\  \cline{2-2} \cline{4-4} \cline{6-6} \cline{8-8} \cline{10-10}
                                                                              & $I=0$           &                                                                                     & {\color{gray}$D_{s}^\pm(2863)^{\dagger}$}  &                                                                                       &  {\color{gray}$D_{s1}^\pm(3214)^{\dagger}$}                &                                                                                       & {\color{gray}$D_{s2}^\pm(3530)^{\dagger}$}          & &     {\color{gray}$D_s^\pm(3538)^{\dagger}$}       \\ \hline
\multirow{3}{*}{\textbf{\begin{tabular}[c]{@{}c@{}}J=L+1\\ S=1\end{tabular}}} & $I=0$           & \multirow{3}{*}{\begin{tabular}[c]{@{}c@{}}100xx3\\ $\mathbf{2^3 S_1}$\\ $1^{--}$\end{tabular}} & $\psi$(2S)   & \multirow{3}{*}{\begin{tabular}[c]{@{}c@{}}100xx5\\ $\mathbf{2^3 P_2}$\\ $2^{++}$\end{tabular}}   & $\chi_{c2}$(3930)            & \multirow{3}{*}{\begin{tabular}[c]{@{}c@{}}100xx7\\ $\mathbf{2^3 D_3}$\\ $3^{--}$\end{tabular}}   & {\color{gray}$\psi_{3}(2D)(4160)^{\dagger}$}       & \multirow{3}{*}{\begin{tabular}[c]{@{}c@{}}200xx3\\ $\mathbf{3^3 S_1}$\\ $1^{--}$\end{tabular}} &  $\psi$(4040)     \\ \cline{2-2} \cline{4-4} \cline{6-6} \cline{8-8} \cline{10-10}
                                                                              & $I=\frac{1}{2}$ &                                                                                     & {\color{gray}$D_1^*(2730)^{\dagger}$}    &                                                                                       & {\color{gray}$D_{2}^*(3052)^{\dagger}$}        &                                                                                       & {\color{gray}${D_{3}^{*\pm}}(3342)^{\dagger}$}      & &   {\color{gray}${D_1^*}(3297)^{\dagger}$}       \\  \cline{2-2} \cline{4-4} \cline{6-6} \cline{8-8}  \cline{10-10}
                                                                              & $I=0$           &                                                                                     & $D_{s1}^{*\pm}$(2700) &                                                                                       & {\color{gray}$D_{s2}^{*\pm}(3025)^{\dagger}$}          &                                                                                       & {\color{gray}$D_{s3}^\pm(3318)^{\dagger}$}     & &   {\color{gray}$D_{s1}^{*\pm}(3181)^{\dagger}$}      \\ \hline
\multirow{3}{*}{\textbf{\begin{tabular}[c]{@{}c@{}}J=L\\ S=1\end{tabular}}}   & $I=0$           & \multicolumn{2}{c|}{\multirow{8}{*}{}}                                                              & \multirow{3}{*}{\begin{tabular}[c]{@{}c@{}}120xx3\\ $\mathbf{2^3 P_1}$\\ $1^{++}$\end{tabular}} &  $\chi_{c1}$(3860)             & \multirow{3}{*}{\begin{tabular}[c]{@{}c@{}}120xx5\\ $\mathbf{2^3 D_2}$\\ $2^{--}$\end{tabular}} & {\color{gray}${\psi_{2}(2D)}(4127)^{\dagger}$}               &  \multicolumn{2}{c}{}              \\ \cline{2-2} \cline{6-6} \cline{8-8} 
                                                                              & $I=\frac{1}{2}$ & \multicolumn{2}{c|}{}                                                                               &                                                                                       &  {\color{gray}${D_{1}}(2913)^{\dagger}$} &                                                                                       & {\color{gray}${D_{2}}(3258)^{\dagger}$}   & \multicolumn{2}{c}{}       \\  \cline{2-2} \cline{6-6} \cline{8-8} 
                                                                              & $I=0$           & \multicolumn{2}{c|}{}                                                                               &                                                                                       &  {\color{gray}${D_{s1}^\pm}(2938)^{\dagger}$}           &                                                                                       &  {\color{gray}${D_{s2}^\pm}(3281)^{\dagger}$}                     & \multicolumn{2}{c}{}    \\ \cline{1-2} \cline{5-8} 
\multirow{3}{*}{\textbf{\begin{tabular}[c]{@{}c@{}}J=L-1\\ S=1\end{tabular}}} & $I=0$           & \multicolumn{2}{c|}{}                                                                               & \multirow{3}{*}{\begin{tabular}[c]{@{}c@{}}110xx1\\ $\mathbf{2^3 P_0}$\\ $0^{++}$\end{tabular}} & $\chi_{c0}$(3860)       & \multirow{3}{*}{\begin{tabular}[c]{@{}c@{}}130xx3\\ $\mathbf{2^3 D_1}$\\ $1^{--}$\end{tabular}} &  $\psi(4160)$         & \multicolumn{2}{c}{}          \\ 
\cline{2-2} \cline{6-6} \cline{8-8}  
                                                                              & $I=\frac{1}{2}$ &         \multicolumn{2}{c|}{}  &   &  {\color{gray}${D_{0}^*}(2921)^{\dagger}$}         &                                                                                       &  {\color{gray}${D_{1}^*}(3308)^{\dagger}$} &  \multicolumn{2}{c}{}   \\ \cline{2-2} \cline{6-6} \cline{8-8} 
                                                                              & $I=0$           &          \multicolumn{2}{c|}{}                                                                         &                                                                                       &  {\color{gray}${D_{s0}^{*\pm}(2935)}^{\dagger}$}            &                                                                                       &  {\color{gray}${D_{s1}^{*\pm}}(3319)^{\dagger}$}      & \multicolumn{2}{c}{}                       \\ \cline{1-2} \cline{5-8} 
\end{tabular}
}
\caption{\label{tab:cmesonradial} Same as Tab.\ \ref {tab:lightmesonradial} but for mesons with charm quarks in radially excited states.
}
\end{table*}
%\endgroup

%\begingroup
\renewcommand{\arraystretch}{1.25}
\begin{table*}[]
{\centering
\begin{tabular}{|c|c||c|c||c|c||c|c||c|c||c|c|}
\hline
\multicolumn{12}{|c|}{\bf Bottom $\mathbf n=1$ ($\mathbf k=0$)}         \\ \hline
\multicolumn{2}{|c||}{}                                                                     & \multicolumn{2}{c||}{\textbf{L=0}}                                                                    & \multicolumn{2}{c||}{\textbf{L=1}}                                                                           & \multicolumn{2}{c||}{\textbf{L=2}}                                                                        & \multicolumn{2}{c||}{\textbf{L=3}}   & \multicolumn{2}{c|}{\textbf{L=4}}                                                                                                                                   \\ \hline\hline
\multirow{4}{*}{\textbf{\begin{tabular}[c]{@{}c@{}}L=J\\ S=0\end{tabular}}}   & $I=0$           & \multirow{4}{*}{\begin{tabular}[c]{@{}c@{}}xx1\\ $\mathbf{1^1 S_0}$\\ $0^{-+}$\end{tabular}} & $\eta_b(1S)$   & \multirow{4}{*}{\begin{tabular}[c]{@{}c@{}}10xx3\\ $\mathbf{1^1 P_1}$\\ $1^{+-}$\end{tabular}} & $h_b(1P)$           & \multirow{4}{*}{\begin{tabular}[c]{@{}c@{}}10xx5\\ $\mathbf{1^1 D_2}$\\ $2^{-+}$\end{tabular}} &  {\color{gray}$\eta_b(1D)(10375)^{\dagger}$}               & \multirow{4}{*}{\begin{tabular}[c]{@{}c@{}}10xx7\\ $\mathbf{1^1 F_3}$\\ $3^{+-}$\end{tabular}} &  {\color{gray}$h_b(1F)(10830)^{\dagger}$}  & \multirow{3}{*}{\begin{tabular}[c]{@{}c@{}}10xx9\\ $\mathbf{1^1 G_3}$\\ $4^{-+}$\end{tabular}} &   {\color{gray}$\eta_b(1G)(11267)^{\dagger}$} \\ \cline{2-2} \cline{4-4} \cline{6-6} \cline{8-8} \cline{10-10} \cline{12-12}
                                                                              & $I=\frac{1}{2}$ &                                                                                     & $B$            &                                                                                       & $B_1$(5721)         &                                                                                       &  {\color{gray}$B_2(6131)^{\dagger}$}                &                                                                                       & {\color{gray}$B_3(6516)^{\dagger}$} & & {\color{gray}$B_4(6879)^{\dagger}$}  \\ \cline{2-2} \cline{4-4} \cline{6-6} \cline{8-8} \cline{10-10} \cline{12-12} 
                                                                              & $I=0$           &                                                                                     & $B_s^0$        &                                                                                       & $B_{s1}^0$(5830)    &                                                                                       &  {\color{gray}$B_{s2}^{0}(6259)^{\dagger}$}               &                                                                                       & {\color{gray}$B_{s3}^{0}(6660)^{\dagger}$}  & & {\color{gray}$B_{s4}^{0}(7039)^{\dagger}$} \\ \cline{2-2} \cline{4-4} \cline{6-6} \cline{8-8} \cline{10-10}  \cline{12-12}
                                                                              & $I=0$           &                                                                                     & $B_c^\pm$      &                                                                                       & {\color{gray}$B_{c1}^{\pm}(6778)^{\dagger}$}                 &                                                                                       & {\color{gray}$B_{c2}^{\pm}(7246)^{\dagger}$}              &                                                                                       & {\color{gray}$B_{c3}^{\pm}(7686)^{\dagger}$}  & & {\color{gray}$B_{c4}^{\pm}(8102)^{\dagger}$} \\ \hline
\multirow{4}{*}{\textbf{\begin{tabular}[c]{@{}c@{}}J=L+1\\ S=1\end{tabular}}} & $I=0$           & \multirow{4}{*}{\begin{tabular}[c]{@{}c@{}}xx3\\ $\mathbf{1^3 S_1}$\\ $1^{--}$\end{tabular}} & $\Upsilon(1S)$ & \multirow{4}{*}{\begin{tabular}[c]{@{}c@{}}xx5\\ $\mathbf{1^3 P_2}$\\ $2^{++}$\end{tabular}}   & $\chi_{b2}(1P)$     & \multirow{4}{*}{\begin{tabular}[c]{@{}c@{}}xx7\\ $\mathbf{1^3 D_3}$\\ $3^{--}$\end{tabular}}   &  {\color{gray}$\Upsilon_3(1D)(10344)^{\dagger}$}              & \multirow{4}{*}{\begin{tabular}[c]{@{}c@{}}xx9\\ $\mathbf{1^3 F_4}$\\ $4^{++}$\end{tabular}}   &  {\color{gray}$\chi_{b4}(1F)(10759)^{\dagger}$} &  \multirow{3}{*}{\begin{tabular}[c]{@{}c@{}} xx8$^\ddagger$ \\ $\mathbf{1^3 G_5}$\\ $5^{--}$\end{tabular}}  &  {\color{gray}$\Upsilon_{5}(1G)(11159)^{\dagger}$} \\ \cline{2-2} \cline{4-4} \cline{6-6} \cline{8-8} \cline{10-10} \cline{12-12}
                                                                              & $I=\frac{1}{2}$ &                                                                                     & $B^*$          &                                                                                       & $B_2^*$(5747)       &                                                                                       &{\color{gray}$B_{3}^{*}(6144)^{\dagger}$}              &                                                                                       & {\color{gray}$B_{4}^{*}(6518)^{\dagger}$} & & {\color{gray}$B_{5}^{*}(6871)^{\dagger}$} \\ \cline{2-2} \cline{4-4} \cline{6-6} \cline{8-8} \cline{10-10} \cline{12-12}
                                                                              & $I=0$           &                                                                                     & $B_s^{*0}$        &                                                                                       & $B_{s2}^{*0}$(5840) &                                                                                       & {\color{gray}$B_{s3}^{*0}(6236)^{\dagger}$}              &                                                                                       &  {\color{gray}$B_{s4}^{*0}(6609)^{\dagger}$}  & &  {\color{gray}$B_{s5}^{*0}(6961)^{\dagger}$}  \\ \cline{2-2} \cline{4-4} \cline{6-6} \cline{8-8} \cline{10-10} \cline{12-12}
                                                                              & $I=0$           &                                                                                     & {\color{gray}$B_c^{*\pm}(6275)^{\dagger}$}            &                                                                                       &  {\color{gray}$B_{c2}^{*\pm}(6743)^{\dagger}$}                  &                                                                                       &  {\color{gray}$B_{c3}^{*\pm}(7180)^{\dagger}$}               &                                                                                       & {\color{gray}$B_{c4}^{*\pm}(7592)^{\dagger}$}   & & {\color{gray}$B_{c5}^{*\pm}(7983)^{\dagger}$}   \\ \hline
\multirow{4}{*}{\textbf{\begin{tabular}[c]{@{}c@{}}J=L\\ S=1\end{tabular}}}   & $I=0$           & \multicolumn{2}{c|}{\multirow{8}{*}{}}                                                               & \multirow{4}{*}{\begin{tabular}[c]{@{}c@{}}20xx3\\ $\mathbf{1^3 P_1}$\\ $1^{++}$\end{tabular}} & $\chi_{b1}(1P)$     & \multirow{4}{*}{\begin{tabular}[c]{@{}c@{}}20xx5\\ $\mathbf{1^3 D_2}$\\ $2^{--}$\end{tabular}} & $\Upsilon_2(1D)$ & \multirow{4}{*}{\begin{tabular}[c]{@{}c@{}}20xx7\\ $\mathbf{1^3 F_3}$\\ $3^{++}$\end{tabular}} &  {\color{gray}$\chi_{b3}(1F)(10428)^{\dagger}$}  & 
\multirow{3}{*}{\begin{tabular}[c]{@{}c@{}} 20xx9 \\ $\mathbf{1^3 G_4}$\\ $4^{--}$\end{tabular}} & {\color{gray}$\Upsilon_{4}(1G)(10685)^{\dagger}$} \\ \cline{2-2} \cline{6-6} \cline{8-8} \cline{10-10} \cline{12-12}
                                                   
                                                                              & $I=\frac{1}{2}$ & \multicolumn{2}{c|}{}                                                                                &                                                                                       & {\color{gray}$B_1(5721)^{\dagger}$}                 &                                                                                       & {\color{gray}$B_2(6178)^{\dagger}$}              &                                                                                       & {\color{gray}$B_3(6603)^{\dagger}$} & & {\color{gray}$B_4(7003)^{\dagger}$} \\ \cline{2-2} \cline{6-6} \cline{8-8} \cline{10-10} \cline{12-12}
                                                   
                                                                              & $I=0$           & \multicolumn{2}{c|}{}                                                                                &                                                                                       & {\color{gray}$B_{s1}^{0}(5830)^{\dagger}$}                   &                                                                                       & {\color{gray}$B_{s2}^{0}(6279)^{\dagger}$}                &                                                                                       & {\color{gray}$B_{s3}^{0}(6698)^{\dagger}$}   & & {\color{gray}$B_{s2}^{0}(7092)^{\dagger}$}   \\ \cline{2-2} \cline{6-6} \cline{8-8} \cline{10-10} \cline{12-12}
                                                   
                                                                              & $I=0$           & \multicolumn{2}{c|}{}                                                                                &                                                                                       &  {\color{gray}$B_{c1}^{*\pm}(6778)^{\dagger}$}                  &                                                                                       & {\color{gray}$B_{c2}^{*\pm}(7168)^{\dagger}$}                &                                                                                       & {\color{gray}$B_{c3}^{*\pm}(7537)^{\dagger}$}   & & {\color{gray}$B_{c4}^{*\pm}(7890)^{\dagger}$}   \\ \cline{1-2} \cline{5-12} 
\multirow{4}{*}{\textbf{\begin{tabular}[c]{@{}c@{}}J=L-1\\ S=1\end{tabular}}} & $I=0$           & \multicolumn{2}{c|}{}                                                                                & \multirow{4}{*}{\begin{tabular}[c]{@{}c@{}}10xx1\\ $\mathbf{1^3 P_0}$\\ $0^{++}$\end{tabular}} & $\chi_{b0}(1P)$     & \multirow{4}{*}{\begin{tabular}[c]{@{}c@{}}30xx3\\ $\mathbf{1^3 D_1}$\\ $1^{--}$\end{tabular}} &  {\color{gray}$\Upsilon_1(1D)(10131)^{\dagger}$}              & \multirow{4}{*}{\begin{tabular}[c]{@{}c@{}}30xx5\\ $\mathbf{1^3 F_2}$\\ $2^{++}$\end{tabular}} &  {\color{gray}$\chi_{b2}(1F)(10396)^{\dagger}$}  & \multirow{3}{*}{\begin{tabular}[c]{@{}c@{}}30xx7 \\ $\mathbf{1^3 G_3}$\\ $3^{--}$\end{tabular}} & {\color{gray}$\Upsilon_{3}(1G)(10654)^{\dagger}$}\\ \cline{2-2} \cline{6-6} \cline{8-8} \cline{10-10} \cline{12-12}
                                                   
                                                                              & $I=\frac{1}{2}$ & \multicolumn{2}{c|}{}                                                                                &                                                                                       & {\color{gray}$B_{0}^{*0}(5721)^{\dagger}$}                 &                                                                                       & {\color{gray}$B_{1}^{*0}(6178)^{\dagger}$}              &                                                                                       & {\color{gray}$B_{2}^{*0}(6603)^{\dagger}$} & & {\color{gray}$B_{3}^{*0}(7003)^{\dagger}$} \\ \cline{2-2} \cline{6-6} \cline{8-8} \cline{10-10} \cline{12-12}
                                                   
                                                                              & $I=0$           & \multicolumn{2}{c|}{}                                                                                &                                                                                       & {\color{gray}$B_{s0}^{*0}(5830)^{\dagger}$}               &                                                                                       & {\color{gray}$B_{s1}^{*0}(6279)^{\dagger}$}              &                                                                                       & {\color{gray}$B_{s2}^{*0}(6698)^{\dagger}$} & & {\color{gray}$B_{s3}^{*0}(7092)^{\dagger}$} \\ \cline{2-2} \cline{6-6} \cline{8-8} \cline{10-10} \cline{12-12}
                                                   
                                                                              & $I=0$           & \multicolumn{2}{c|}{}                                                                                &                                                                                       & {\color{gray}$B_{c0}^{*\pm}(6778)^{\dagger}$}                   &                                                                                       & {\color{gray}$B_{c1}^{*\pm}(7168)^{\dagger}$}                &                                                                                       & {\color{gray}$B_{c2}^{*\pm}(7537)^{\dagger}$}   & & {\color{gray}$B_{c3}^{*\pm}(7890)^{\dagger}$}  \\ \cline{1-2} \cline{5-12} 
\end{tabular}}
\caption{\label{tab:bmeson} Same as Tab.\ \ref{tab:cmeson} for bound states of bottom quarks with light and charm quarks. Two isospin $I=1/2$ (bottom+up/down) and three different $I=0$ states ($b\bar b$, bottom+strange, bottom+charm) are possible.}
\end{table*}
%\endgroup

%\begingroup
\renewcommand{\arraystretch}{1.25}
\begin{table*}[]
{\centering
\begin{tabular}{|c||c|c|c|c|c|c|c||c|c|}
\hline
\multicolumn{8}{|c||}{\bf Bottom $\mathbf n=2$ ($\mathbf k=1$)}  & \multicolumn{2}{c|}{\bf $\mathbf n=3$ ($\mathbf k=2$)}       \\ \hline
\multicolumn{2}{|c|}{}                                                              & \multicolumn{2}{c|}{\textbf{L=0}}                                                                   & \multicolumn{2}{c|}{\textbf{L=1}}                                                                              & \multicolumn{2}{c||}{\textbf{L=2}}                                                                                                                                                 
 & \multicolumn{2}{c|}{\textbf{L=0}}                                                                                                                                                
 \\ \hline
\multirow{3}{*}{\textbf{\begin{tabular}[c]{@{}c@{}}L=J\\ S=0\end{tabular}}}   & $I=0$           & \multirow{3}{*}{\begin{tabular}[c]{@{}c@{}}100xx1\\ $\mathbf{2^1 S_0}$\\ $0^{-+}$\end{tabular}} & $\eta_b$(2S)         & \multirow{3}{*}{\begin{tabular}[c]{@{}c@{}}110xx3\\ $\mathbf{2^1 P_1}$\\ $1^{+-}$\end{tabular}} &  $h_{b}(2P)$           & \multirow{3}{*}{\begin{tabular}[c]{@{}c@{}}110xx5\\ $\mathbf{2^1 D_2}$\\ $2^{-+}$\end{tabular}} &  {\color{gray}$\eta_{b}(2D)(10515)^{\dagger}$}                
& \multirow{3}{*}{\begin{tabular}[c]{@{}c@{}}200xx1\\ $\mathbf{3^1 S_0}$\\ $0^{-+}$\end{tabular}}  & {\color{gray}$\eta_b(3S)(10475)^{\dagger}$}  \\ \cline{2-2} \cline{4-4} \cline{6-6} \cline{8-8} \cline{10-10}
                                                                              & $I=\frac{1}{2}$ &                                                                                     &  {\color{gray}$B(6133)^{\dagger}$}           &                                                                                       &  {\color{gray}$B_1(6550)^{\dagger}$}    &                                                                                       & {\color{gray}$B_{2}(6942)^{\dagger}$}  & & {\color{gray}$B_3(6882)^{\dagger}$}   \\  \cline{2-2} \cline{4-4} \cline{6-6} \cline{8-8} \cline{10-10} 
 & $I=0$ &                                                                                     & {\color{gray}${B_{s}^0(6209)}^{\dagger}$}           &                                                                                       &  {\color{gray}${B_{s1}^0(6621)}^{\dagger}$}    &                                                                                       & {\color{gray}${B_{s2}^0(7009)}^{\dagger}$}  & & {\color{gray}${B_{s}^0(6949)}^{\dagger}$} \\  \cline{2-2} \cline{4-4} \cline{6-6} \cline{8-8} \cline{10-10}
                                                                              & $I=0$           &                                                                                     & $B_{c}(2S)^\pm$  &                                                                                       &   {\color{gray}${B_{c1}^\pm(7247)}^{\dagger}$}             &                                                                                       &  {\color{gray}${B_{c2}^\pm(7603)}^{\dagger}$}          & &      {\color{gray}${B_{c}^\pm(7548)}^{\dagger}$}      \\ \hline
\multirow{3}{*}{\textbf{\begin{tabular}[c]{@{}c@{}}J=L+1\\ S=1\end{tabular}}} & $I=0$           & \multirow{3}{*}{\begin{tabular}[c]{@{}c@{}}100xx3\\ $\mathbf{2^3 S_1}$\\ $1^{--}$\end{tabular}} & $\Upsilon$(2S)   & \multirow{3}{*}{\begin{tabular}[c]{@{}c@{}}100xx5\\ $\mathbf{2^3 P_2}$\\ $2^{++}$\end{tabular}}   & $\chi_{b2}(2P)$            & \multirow{3}{*}{\begin{tabular}[c]{@{}c@{}}100xx7\\ $\mathbf{2^3 D_3}$\\ $3^{--}$\end{tabular}}   & {\color{gray}${\Upsilon_{3}(2D)}(10509)^{\dagger}$}      & \multirow{3}{*}{\begin{tabular}[c]{@{}c@{}}200xx3\\ $\mathbf{3^3 S_1}$\\ $1^{--}$\end{tabular}} &  $\Upsilon(3S)$     \\ \cline{2-2} \cline{4-4} \cline{6-6} \cline{8-8} \cline{10-10}
                                                                              & $I=\frac{1}{2}$ &                                                                                     & {\color{gray}$B^*(6140)^{\dagger}$}    &                                                                                       & {\color{gray}$B_2^*(6534)^{\dagger}$}        &                                                                                       & {\color{gray}$B_3^*(6906)^{\dagger}$}     & &   {\color{gray}$B^*(6863)^{\dagger}$}       \\  \cline{2-2} \cline{4-4} \cline{6-6} \cline{8-8}  \cline{10-10}
& $I=0$ &                                                                                     & {\color{gray}$B_s^*(6223)^{\dagger}$}            &                                                                                       &  {\color{gray}$B_{s2}^*(6612)^{\dagger}$}      &                                                                                       & {\color{gray}$B_{s3}^*(6979)^{\dagger}$} & & {\color{gray}$B_s^*(6937)^{\dagger}$}   \\  \cline{2-2} \cline{4-4} \cline{6-6} \cline{8-8} \cline{10-10}
                                                                              & $I=0$           &                                                                                     & $B_{c}^{*}(2S)^\pm$ &                                                                                       & {\color{gray}$B_{c2}^{*\pm}(7225)^{\dagger}$}          &                                                                                       & {\color{gray}$B_{c3}^{*\pm}(7563)^{\dagger}$}     & &   {\color{gray}$B_{c}^{*\pm}(7524)^{\dagger}$}      \\ \hline
\multirow{3}{*}{\textbf{\begin{tabular}[c]{@{}c@{}}J=L\\ S=1\end{tabular}}}   & $I=0$           & \multicolumn{2}{c|}{\multirow{8}{*}{}}                                                              & \multirow{3}{*}{\begin{tabular}[c]{@{}c@{}}120xx3\\ $\mathbf{2^3 P_1}$\\ $1^{++}$\end{tabular}} &  $\chi_{b1}(2P)$             & \multirow{3}{*}{\begin{tabular}[c]{@{}c@{}}120xx5\\ $\mathbf{2^3 D_2}$\\ $2^{--}$\end{tabular}} & {\color{gray}${\Upsilon_{2}(2D)}(10503)^{\dagger}$}             &  \multicolumn{2}{c}{}              \\ \cline{2-2} \cline{6-6} \cline{8-8} 
                                                                              & $I=\frac{1}{2}$ & \multicolumn{2}{c|}{}                                                                               &                                                                                       &  {\color{gray}${B_{1}}(6326)^{\dagger}$} &                                                                                       & {\color{gray}${B_{2}}(6720)^{\dagger}$}   & \multicolumn{2}{c}{}       \\  \cline{2-2} \cline{6-6} \cline{8-8} 
& $I=0$ & \multicolumn{2}{c|}{}                                                                               &                                                                                       &  {\color{gray}${B_{s1}^0}(6425)^{\dagger}$} &                                                                                       & {\color{gray}${B_{s2}^0}(6813)^{\dagger}$}   & \multicolumn{2}{c}{}       \\  \cline{2-2} \cline{6-6} \cline{8-8} 

                                                                              & $I=0$           & \multicolumn{2}{c|}{}                                                                               &                                                                                       &  {\color{gray}${B_{c1}^\pm}(7296)^{\dagger}$}           &                                                                                       &  {\color{gray}${B_{c2}^\pm}(7640)^{\dagger}$}                     & \multicolumn{2}{c}{}    \\ \cline{1-2} \cline{5-8} 
\multirow{3}{*}{\textbf{\begin{tabular}[c]{@{}c@{}}J=L-1\\ S=1\end{tabular}}} & $I=0$           & \multicolumn{2}{c|}{}                                                                               & \multirow{3}{*}{\begin{tabular}[c]{@{}c@{}}110xx1\\ $\mathbf{2^3 P_0}$\\ $0^{++}$\end{tabular}} & $\chi_{b0}(2P)$       & \multirow{3}{*}{\begin{tabular}[c]{@{}c@{}}130xx3\\ $\mathbf{2^3 D_1}$\\ $1^{--}$\end{tabular}} & {\color{gray}${\Upsilon_{1}(2D)}(10481)^{\dagger}$}        & \multicolumn{2}{c}{}          \\ 
\cline{2-2} \cline{6-6} \cline{8-8}  
                                                                              & $I=\frac{1}{2}$ &         \multicolumn{2}{c|}{}  &   &  {\color{gray}${B_{0}^*}(6344)^{\dagger}$}         &                                                                                       &  {\color{gray}${B_{1}^*}(6737)^{\dagger}$} &  \multicolumn{2}{c}{}   \\ \cline{2-2} \cline{6-6} \cline{8-8} 
& $I=0$ & \multicolumn{2}{c|}{}                                                                               &                                                                                       &  {\color{gray}${B_{s0}^0}(6442)^{\dagger}$} &                                                                                       & {\color{gray}${B_{s1}^0}(6830)^{\dagger}$}   & \multicolumn{2}{c}{}       \\  \cline{2-2} \cline{6-6} \cline{8-8} 
                                                                              & $I=0$           &          \multicolumn{2}{c|}{}                                                                         &                                                                                       &  {\color{gray}${B_{c0}^{*\pm}(7311)}^{\dagger}$}            &                                                                                       &  {\color{gray}${B_{c1}^{*\pm}}(7655)^{\dagger}$}      & \multicolumn{2}{c}{}                       \\ \cline{1-2} \cline{5-8} 
\end{tabular}
}
\caption{\label{tab:bmesonradial} Same as Tab.\ \ref {tab:lightmesonradial} but for mesons with bottom quarks in radially excited states.
}
\end{table*}
%\endgroup

\end{appendix}

\bibliography{ExcitedReco}

\end{document}